\documentclass[a4paper]{article}
\usepackage[a4paper, left = 2cm, right = 2cm, top = 2.5cm, bottom = 3.5cm]{geometry}

\usepackage{booktabs} 
\usepackage{authblk}

\usepackage[utf8x]{inputenc}
\usepackage[english]{babel}   
\usepackage[export]{adjustbox}

\usepackage{graphicx}
\usepackage{subfig}

\usepackage{graphicx} 
\usepackage{hyperref}
\usepackage{amsmath}
\usepackage{amssymb}
\usepackage{slashed}
\usepackage{cite}
\usepackage{placeins}

\newcommand{\nn}{\nonumber}
\newcommand{\be}{\begin{equation}} 
\newcommand{\ee}{\end{equation}}
\newcommand{\bea}{\begin{eqnarray}} 
\newcommand{\eea}{\end{eqnarray}}
\newcommand{\beaV}{\begin{equation}\begin{aligned}} 
\newcommand{\eeaV}{\end{aligned}\end{equation}}

\usepackage{color}

\definecolor{KBFIred}{RGB}{163,35,47}

\begin{document}

\title{ \textcolor{KBFIred}{Simplified leptoquark models for precision $l_i \rightarrow l_f \gamma$ experiments: \\two-loop structure of $O(\alpha_s Y^2)$ corrections}}

\author[a]{Luigi Delle Rose\thanks{luigi.dellerose@fi.infn.it}}
\author[b]{Carlo Marzo\thanks{carlo.marzo@kbfi.ee; (corresponding author)}}
\author[b]{Luca Marzola\thanks{luca.marzola@cern.ch}}

\affil[a]{INFN, Sezione di Firenze, and Department of Physics and Astronomy, University of Florence,

Via G. Sansone 1, 50019 Sesto Fiorentino, Italy}
\affil[b]{NICPB, R\"avala Pst. 10, 10143 Tallinn, Estonia}
\date{}
\maketitle

\begin{abstract} 
\noindent
We put forward the framework of simplified leptoquark models: simple extensions of the Standard Model that serve as benchmarks to test the interactions of leptons with new colored degrees of freedom considered, for instance, in leptoquark or grand unification models. As a first application of the scheme, we analyze the power of precision lepton observables by computing gauge invariant two-loop radiative corrections to the lepton-photon vertex, generated here by Yukawa interactions between the lepton and new colored degrees of freedom. The result, detailed in explicit expressions for the involved form factors, improves on the literature for the higher loop order considered and highlights the existence of regions in the parameter space of the model where two-loop corrections cannot be neglected.   
\vspace{2 cm}
\end{abstract}

%-------------------------------------------------------------------------------
\section{Introduction}
%-------------------------------------------------------------------------------
After more than one decade of LHC data in substantial agreement with the known properties of physics at the electroweak scale, the possibility of finding new light states strongly coupled to the Standard Model (SM) particles seems certainly unlikely. Observations have pushed any possible new physics scale well above the heaviest SM particle, reaching up to the multi-TeV range that will be fully explored by the next-generation collider experiments~\cite{Benedikt:2018csr,Abada:2019zxq,Abada:2019ono,Abada:2019lih}. 

In this context, low energy leptonic observables provide clean and sensitive benchmarks for the most precise predictions of the SM and, consequently, are of great importance for their potential to highlight possible discrepancies between theory and experiment. 

While precision lepton physics is being probed in an array of experiments with an unprecedented accuracy~\cite{Grange:2015fou,PhysRevLett.100.120801,Hanneke:2010au,TheMEG:2016wtm}, on the theoretical side it is thus mandatory to determine the relevant higher order corrections. In particular, the vertex functions for the leptonic transitions $\ell_i \to \ell_f \gamma$, have been the subject of intense investigation that led to an analytical determination of the four-loop QED contribution~\cite{Laporta:2017okg}, a numerical computation of the five-loop one~\cite{Aoyama:2012wj} and the calculation of analytical expressions for the two-loop electroweak corrections~\cite{Czarnecki:1995sz}. 

In some cases, interestingly, the improved estimate of the SM contribution resulted in tensions with the experiments. The paradigmatic case is that of the longstanding anomaly in the muon response to a static magnetic field, usually referred to as the anomalous magnetic moment (AMM) $(g-2)_{\mu}$. Pending the results of the Fermilab E989 experiments~\cite{Grange:2015fou}, the current world average dominated by the Brookhaven experiment E821~\cite{Bennett:2006fi},
\bea
a_{\mu}^{exp} = (116592091\pm 54 \pm 33) \times 10^{-11},
\eea
confronts the SM prediction
\bea
a_{\mu}^{SM} = (116591811\pm 62 ) \times 10^{-11},
\eea  
where hadronic effects dominate the uncertainty. The current discrepancy is therefore 
\bea
\label{eq:g2mu}
a_{\mu}^{exp} - a_{\mu}^{SM} = (278\pm 88) \times 10^{-11},
\eea
amounting to a $3.1\sigma$ difference\footnote{A recent lattice determination of the SM hadronic vacuum polarization contribution~\cite{Borsanyi:2020mff} has put into discussion the significance of the reported anomaly. While the consequences of this computation are still being debated~\cite{Crivellin:2020zul}, we regard the range in Eq.~\eqref{eq:g2mu} as an indication of the sensitivity characterizing the observable.}. 

Interestingly, a new determination of the fine structure constant has revealed a similar anomaly in the electron sector, where a $2.4\sigma$ tension concerns the corresponding anomalous magnetic moment $(g-2)_e$~\cite{PhysRevLett.100.120801,Hanneke:2010au}:
\bea
a_{e}^{exp} - a_{e}^{SM} = (-87\pm 36) \times 10^{-14}.
\eea
It is a puzzling aspect, and a challenging issue at the model-building level, that the two anomalies have different signs. Furthermore, the deviation in the anomalous magnetic moment of the electron is in magnitude larger than the muon one, after the $m_e/m_{\mu}$ rescaling is taken into account~\cite{Davoudiasl:2018fbb,Crivellin2018}.  

Besides AMM, there are other interactions between leptons and a photon which have the potential to yield tight bounds on any new physics model contributing to the full lepton-photon vertex. For instance, sources of CP violations transcending the SM one are highly constrained by the observations of the electric dipole moment (EDM) of the electron, $d_e$, through the limit~\cite{ACME}
\bea
d_{e} <  1.1\times 10^{-29} e \cdot \text{cm}.
\eea
The same parameters will be challenged by future refinements of the muon EDM $d_{\mu}$~\cite{Grozin:2009jq,Crivellin:2018qmi}, currently measured in
\bea \label{dmulim}
d_{\mu} <  0.9\times 10^{-19} e \cdot \text{cm}.
\eea  
Similarly, sources of lepton flavor violation (LFV) beyond the neutrino mixing generally trigger the transition $\mu \rightarrow e \gamma$, strongly constrained by the experimental bound~\cite{TheMEG:2016wtm}
\bea
\textrm{Br}(\mu \rightarrow e \gamma) < 4.2 \times 10^{-13}.
\eea

Motivated by the need for improved computations of precision leptonic observables, in this paper we go beyond the results in the literature by assessing the impact of these bounds at a further loop order. To this purpose  we introduce the framework of simplified leptoquark models (SLMs): effective SM extensions meant as a testing ground for more complex models that contain new colored degrees of freedom coupled to the SM leptons. In this first analysis we estimate the power of precision leptonic observable to constrain scenarios which contain a new scalar field and, at most, an extra fermionic degree of freedom, both partaking in strong and electromagnetic interactions. The complementary case of vector extensions of the SM will instead be addressed in a forthcoming paper. In the following we therefore detail the two-loop structure of the effective $\ell \ell \gamma$ vertex up the order $\mathcal{O}(\alpha_s y^2)$, with $y$ being the Yukawa coupling of the new scalar field, giving the explicit expressions of the form factors that determine the mentioned leptonic observables. 

The relevance of our analysis is manifested by the ubiquity of these colored particles, postulated in many beyond SM theories and motivated, for instance, by the B-physics anomalies or unification scenarios~\cite{Becirevic:2016yqi,Mandal:2018kau,Dorsner:2016wpm,Bandyopadhyay:2018syt,Angelescu:2018tyl,Faber:2018afz,Aebischer:2018acj,Aydemir:2019ynb,Popov:2019tyc,Blanke:2019aao,Kosnik:2019axo,Crivellin:2019dwb,Borschensky:2020hot,Bigaran:2020jil,Arnan:2019olv}. In regard of this, the $S_1$ and $R_2$ classes of leptoquark models can be straightforwardly studied by using the SLMs. On the theoretical level, instead, our analysis is influenced by the seminal work in Ref.~\cite{Barr:1990vd}, followed by the investigations in Refs.~\cite{Chang:1990sf,Chang:2000ii,Arhrib:2001xx,Ilisie:2015tra,Abe:2013qla}, which assess the relative relevance of the one and two-loop contributions. In fact, on top of the possible presence of logarithmic and mass-ratio enhancements at next-to-leading order (NLO), the effects of strong interactions easily top other contributions present at the same loop order, thereby justifying the adopted framework.

The paper is structured as follows: in Sec.~\ref{proj} we present the form factors and the corresponding projectors, considering both flavor-conserving and flavor-violating lepton-photon interactions. The scalar SLMs are introduced in Sec.~\ref{scm}, where we detail our computational method and give analytical expressions for the form factors in three scenarios delineated by different mass hierarchies. The results obtained are presented in Sec.~\ref{plots}, whereas in Sec.~\ref{sec:RK} we comment on the complementary bounds due to the $Z$ boson phenomenology. Our work is summarized in Sec.~\ref{conclusions}.

%-------------------------------------------------------------------------------
\section{Form factors and projection operators}\label{proj} 
\begin{figure}[h]
	\centering
	\includegraphics[width=.2\textwidth]{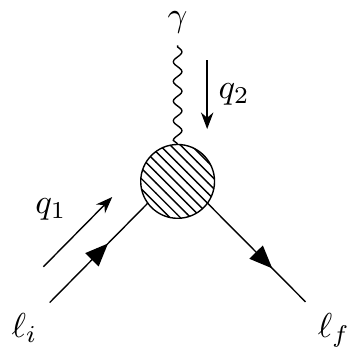}
	\caption{The effective vertex under scrutiny. The label $\ell_{i,f}$ stand for the initial and final lepton, respectively.}
	\label{ampGen1}
\end{figure}

%-------------------------------------------------------------------------------
We begin by detailing the form factors that enter the analyzed effective vertex, shown in Fig.~\ref{ampGen1}, for the cases of lepton flavor-violating  ($\ell_i\neq\ell_f$) and lepton flavor-conserving ($\ell_i=\ell_f$) transitions. 
\subsection{Flavor conserving transitions}
The matrix element for flavour conserving transitions $\ell \rightarrow \ell \gamma$ can be parametrized as follows\footnote{We adopt the same conventions for the Dirac algebra and traces as those used in {\tt FORM}~\cite{Ruijl:2017dtg}.}
\bea \label{amp}
\langle \ell |M_{\mu}| \ell \rangle &=& e \cdot \bar{u}(q_1+q_2)\left[F_1(t)\gamma_{\mu} - \frac{i}{2 m_\ell}F_2(t)\sigma_{\mu \nu} q_2^{\nu} + \frac{1}{m_\ell}F_3(t)q_2{}_{\mu} + \right. \nn \\ && \left. +\gamma_5 \left( G_1(t)\gamma_{\mu} - \frac{i}{2 m_\ell}G_2(t)\sigma_{\mu \nu} q_2^{\nu} + \frac{1}{m_\ell}G_3(t)q_2{}_{\mu}\right) \right] u(q_1) ,
\eea
where $q_2^2 = t$ and $q_1^2 = (q_1 + q_2)^2 = m_\ell^2$. In Eq.~\eqref{amp} $F_2(t)$ and $G_2(t)$ are, respectively, the magnetic and the electric form factor. Within renormalizable theories these quantities are necessarily finite, being the coefficients of dimension 5 operators. Observations are matched in the soft photon limit, $t\to 0$, where we recover the AMM $a_\ell = F_2(0)$ and the EDM $d_\ell = i\, G_2(0)/(2 m_\ell)$. As for the remaining terms, $F_1(t)$ is the charge form factor, whereas $F_3(t)$ vanishes because of the electromagnetic current conservation. 

The expression for the individual form factors can be obtained through the projection operator
\bea 
\label{projlc}
P_{\mu} &=& (\slashed{q}_1 + m_\ell)\left\{ \left[f_1 \gamma_{\mu} - \frac{f_2}{m_\ell} \left(q_1{}_{\mu} + \frac{q_2{}_{\mu}}{2}\right) - \frac{f_3}{m_\ell} q_2{}_{\mu} \right] + \right. \nn \\ && \left. + \gamma_5 \left[g_1 \gamma_{\mu} - \frac{g_2}{m_\ell} \left(q_1{}_{\mu} + \frac{q_2{}_{\mu}}{2}\right) - \frac{g_3}{m_\ell} q_2{}_{\mu} \right] \right\} (\slashed{q}_1 + \slashed{q}_2 + m_\ell)\,,
\eea 
by setting the coefficients $f_i$ and $g_i$ to satisfy the six independent conditions $Tr[P_{\mu} M^{\mu}] = F_i$ and $Tr[P_{\mu} M^{\mu}] = G_i$. In the first two lines of Tab.~\ref{tab:proj} we report the values required to isolate the expressions of $F_2(t)$ and $G_2(t)$ used in the following analysis.

\subsection{Flavor violating transitions}
In considering flavor violating amplitudes $\ell_i \rightarrow \ell_f \gamma$ we take the final state lepton to have negligible mass, for the sake of simplicity. The matrix element then acquires the form
\bea \label{ampmutoe}
\langle \ell_f |M_{\mu}| \ell_i \rangle &=& e \cdot \bar{u}_f(q_1+q_2)\left[\tilde{F}_1(t)\gamma_{\mu} - \frac{i}{2 m_{\ell_i}}\tilde{F}_2(t)\sigma_{\mu \nu} q_2^{\nu} + \frac{1}{m_{\ell_i}}\tilde{F}_3(t)q_2{}_{\mu} + \right. \nn \\ && \left. +\gamma_5 \left( \tilde{G}_1(t)\gamma_{\mu} - \frac{i}{2 m_{\ell_i}}\tilde{G}_2(t)\sigma_{\mu \nu} q_2^{\nu} + \frac{1}{m_{\ell_i}}\tilde{G}_3(t)q_2{}_{\mu}\right) \right] u_i(q_1) \,, 
\eea
whereas the generic form of the projector is: 
\bea 
\label{projlv}
P_{\mu} &=& (\slashed{q}_1 + m_{\ell_i})\left\{ \left[f_1 \gamma_{\mu} - \frac{f_2}{m_\ell} \left(q_1{}_{\mu} + \frac{q_2{}_{\mu}}{2}\right) - \frac{f_3}{m_{\ell_i}} q_2{}_{\mu} \right] + \right. \nn \\ && \left. + \gamma_5 \left[g_1 \gamma_{\mu} - \frac{g_2}{m_{\ell_i}} \left(q_1{}_{\mu} + \frac{q_2{}_{\mu}}{2}\right) - \frac{g_3}{m_{\ell_i}} q_2{}_{\mu} \right] \right\} (\slashed{q}_1 + \slashed{q}_2)\,.
\eea 
The form factors relevant for the proposed analyses can be isolated by using the values of the six $f_i$ and $g_i$ constants reported in last lines of Tab.~\ref{tab:proj}.

\begin{table}[htb!]
	\centering
	\setlength\tabcolsep{4.5pt}
	\begin{tabular}{p{0.9cm}|c|c|c|c|c|c}
		\toprule
		Form factor  & $f_1$  & $f_2$& $f_3$& $g_1$  &$g_2$  &$g_3$\\
		\midrule
		$F_2(t)$ & $\frac{2 m_\ell^2}{(d-2)t(t-4 m_\ell^2)}$  & $-\frac{2 m_\ell^2(4 m_\ell^2 + (d-2)t)}{(d-2)t(t-4 m_\ell^2)^2}$ & 0 & 0 & 0 & 0\\[3ex]
	   $G_2(t)$ & 0 & 0 & 0 & 0 & $\frac{2 m_\ell^2}{t(t - 4 m_\ell^2)}$& 0\\[3ex]
	   \midrule
	   $\tilde F_2(t)$ & $\frac{m_{\ell_i}^2}{2(d-2)(m_{\ell_i}^2-t)^2}$  & $\frac{m_{\ell_i}^2 (m_{\ell_i}^2 + t (d-2))}{2(d-2)(m_{\ell_i}^2-t)^3}$ & $\frac{m_{\ell_i}^4 (d-1)}{4(d-2)(m_{\ell_i}^2-t)^3}$ & 0 & 0 & 0\\[3ex]
	   $\tilde G_2(t)$ & 0 & 0 & 0 & $\frac{m_{\ell_i}^2}{2(d-2)(m_{\ell_i}^2-t)^2}$ & $- \frac{m_{\ell_i}^2 (m_{\ell_i}^2 + t (d-2))}{2(d-2)(m_{\ell_i}^2-t)^3} $& $\frac{m_{\ell_i}^4 (d-1)}{4(d-2)(m_{\ell_i}^2-t)^3}$\\[3ex]
		\bottomrule
	\end{tabular}
	\caption{Values of the $f_i$ and $g_i$ coefficients, $i=1,2,3$, used to isolate the indicated form factors from Eq.~\eqref{amp} (or~\eqref{ampmutoe}) via Eq.~\eqref{projlc} (or~\eqref{projlv}) for lepton flavor conserving (or violating; indicated with a tilde) transitions. In all expressions $d$ stands for the number of spacetime dimensions. }
	\label{tab:proj}
\end{table}

%-------------------------------------------------------------------------------
\section{Simplified leptoquark models} \label{scm}
%-------------------------------------------------------------------------------

As a first example of SLMs we consider extensions of the SM that contain one scalar field, taken in the (anti-) fundamental representation of $SU(3)_c$, and, at most, one new fermionic degree of freedom to account for gauge invariance\footnote{For the purposes of our computation, the sign change that appears in the $guu$ interaction vertex, depending on whether $u$ transforms according to the fundamental or anti-fundamental representation of $SU(3)c$,  is always compensated by the sign in the corresponding vertex involving $\Phi$. Therefore, assigning $u$ or $\Phi$ to the fundamental representation of $SU(3)_c$ makes no difference at the level of the considered observables.}. In the spirit of simplified models, the framework disregards the additional degrees of freedom implied by weak interactions. In fact, whereas complete models that embed the proposed degree of freedoms need to address these interactions in full, the proposed scheme is certainly enough to gauge the dominant contributions of new physics into leptonic precision observables. The Lagrangian at hand then is     
\bea \label{yuk}
\mathcal{L}_{Y} = \mathcal{L}_{0}(\Phi, u)+ \bar{u} ( P_L Y_{L_i} + P_R Y_{R_i} ) \ell_i \, \Phi^{\dagger} + \bar{\ell}_f ( P_L Y_{R_f}^{\dagger} + P_R Y_{L_f}^{\dagger} )u \, \Phi \quad (+ \text{ H.c. interactions if $i \neq f$}) \, ,
\eea
where $\mathcal{L}_{0}(\Phi, u)$ contains the kinetic terms of the new degrees of freedom and $\ell_{i,f}$ are SM charged leptons. The $U(1)_{QED}$ charges of the new fields are left free but must obey the charge conservation condition $Q_{\Phi} = - 1 - Q_{u}$. 

The interactions in Eq.~\eqref{yuk} allow us to model the dominant $\mathcal{O}(\alpha_s Y^2)$ contribution of new physics into the vertex of Fig.~\ref{ampGen1}, recovering for instance the case of scalar leptoquark scenarios. In fact, when $u$ is identified with the (Majorana conjugated) top quark, Eq.~\eqref{yuk} results from the Lagrangian of the $S_1 \sim (3, 1, −1/3)$ or $R_2 \sim (3, 2, 7/6)$ scalar leptoquark models, after the SM spontaneous symmetry breaking. A full embedding of the present model into these framework is then straightforwardly obtained by assigning the degrees of freedom considered here to the appropriate $SU(2)$ multiplets, as well as by setting the corresponding hypercharges to the desired values. 
 
Differently, if $u$ is a new fermionic field, we can explore a larger set of models bounded only by the requirement that $u$ be heavier than the involved SM lepton, as imposed by our methodology. In regard of this, we will focus on three different scenarios characterized by complementary mass hierarchies  
\begin{itemize}
\item Scenario I: $M_{u} \gg M_\Phi \gg m_\ell$
\item Scenario II: $M_{\Phi} \sim M_u\gg m_\ell$
\item Scenario III: $M_{\Phi} \gg M_u\gg m_\ell$ 
\end{itemize}
which require different sub-diagram expansions. In the next sections we first clarify the meaning of this last statement by laying out the employed  methodology. Afterwards, we detail the leading order (LO) and NLO contributions to the form factors $F_2(0)$ and $G_2(0)$ introduced in Sec.~\ref{proj}. 

%-------------------------------------------------------------------------------
\subsection{Methodology}
%-------------------------------------------------------------------------------

Loop computations greatly benefit from the presence of sharp scale hierarchies, which encourage the use of natural series expansions to simplify the integrations involved. However, the result of a series expansion and loop integration is generally sensitive to the order in which these operations are taken and, consequently, more elaborate methods must be employed to address this issue. The rules for a correct asymptotic expansion around a large mass limit have been elaborated in Refs.~\cite{Smirnov:1990rz,Smirnov:1994tg,Smirnov:2002pj,Fleischer:1998nc} in the form of an expansion in sub-diagrams defined as 
\bea
F_G(q,M,m,\epsilon) \xrightarrow{M\rightarrow \infty} \sum_{\sigma} F_{G/\sigma}(q,M,m,\epsilon) \cdot T_{q^{\sigma},m^{\sigma}} F_{\sigma}(q^{\sigma},M,m^{\sigma},\epsilon).
\eea
In the formula above, $T$ is the Taylor expansion operator that acts on the Feynman integral $F_{\sigma}$ corresponding to the sub-diagram $\sigma$, which depends on the mass $m^\sigma$ and momentum $q^\sigma$, while $F_{G/\sigma}$ is the Feynman integral for the original diagram $G$ after the sub-diagram $\sigma$ has been collapsed to a simple vertex. The summation is on all sub-diagrams $\sigma$ that contain all lines associated with the large mass $M$ and, also, are one-particle irreducible with respect to the lines associated to the small mass $m$. 

\begin{figure}[h]
	\centering
	\includegraphics[width=.25\linewidth]{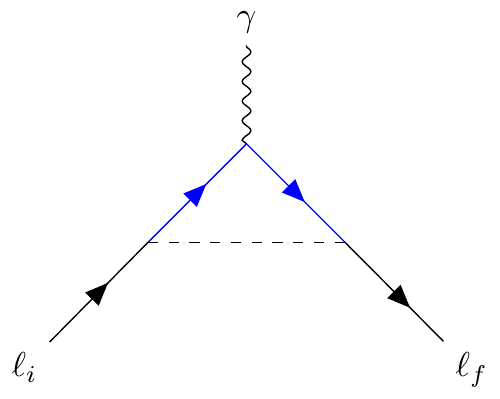}
	\hspace{1cm}
	\includegraphics[width=.25\linewidth]{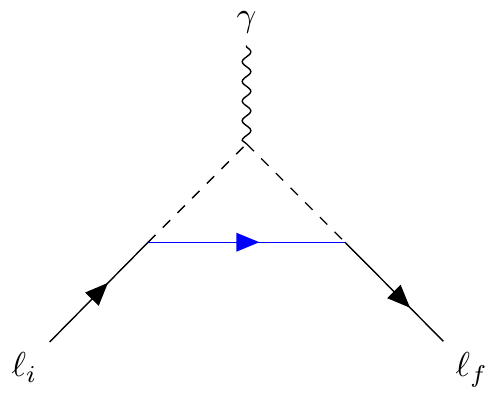}
		\caption{The two contributions into the analyzed lepton-lepton-photon vertex sourced by Eq.~\eqref{yuk} at the one-loop level. The blue fermionic lines are associated to the $u$ field.}
	\label{1LQEDLQ}
\end{figure}

\begin{figure}[h]
	\centering
	\includegraphics[width=.195\textwidth]{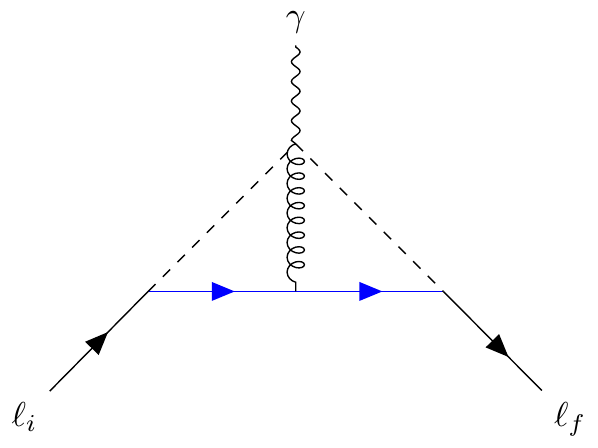} 
	\includegraphics[width=.195\textwidth]{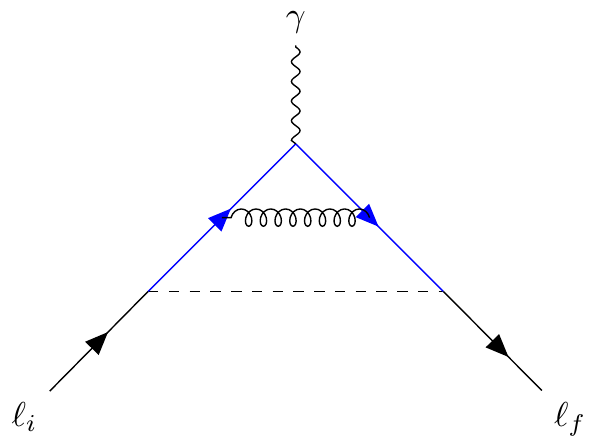}
	\includegraphics[width=.195\textwidth]{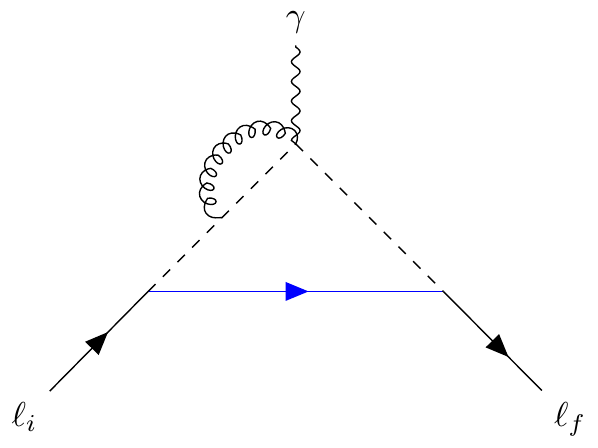}
	\includegraphics[width=.195\textwidth]{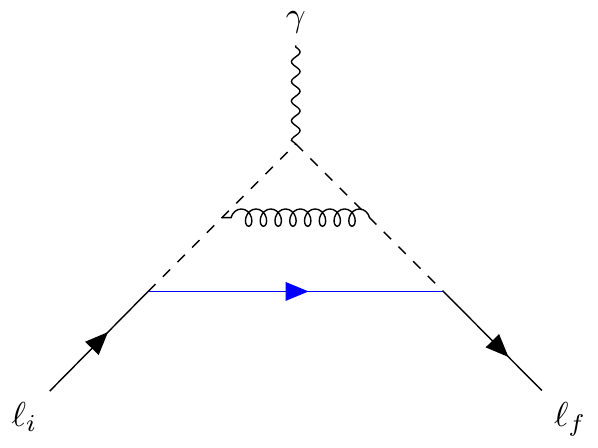}
	\includegraphics[width=.195\textwidth]{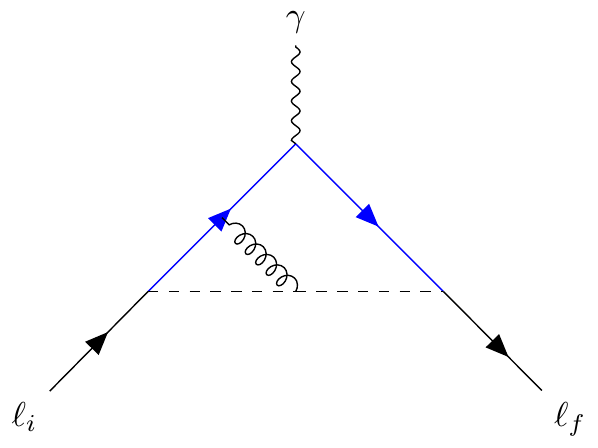}\\
	\includegraphics[width=.195\textwidth]{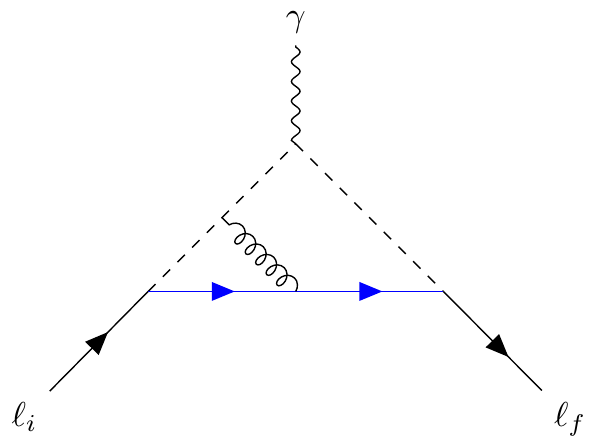}
	\includegraphics[width=.195\textwidth]{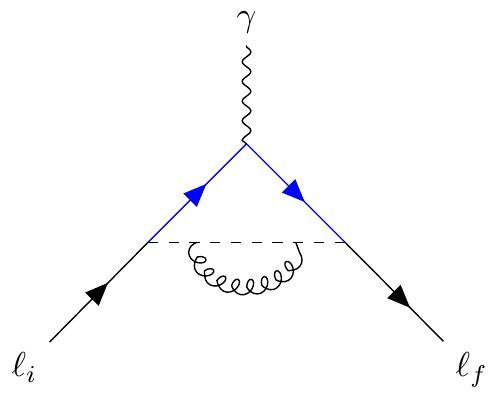}
	\includegraphics[width=.195\textwidth]{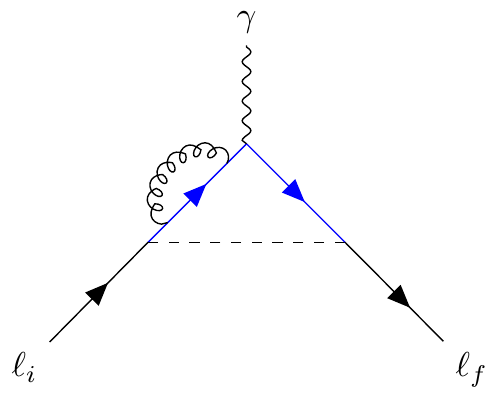}
	\includegraphics[width=.195\textwidth]{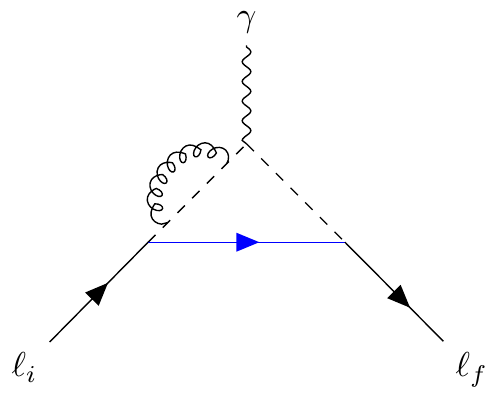}
	\includegraphics[width=.195\textwidth]{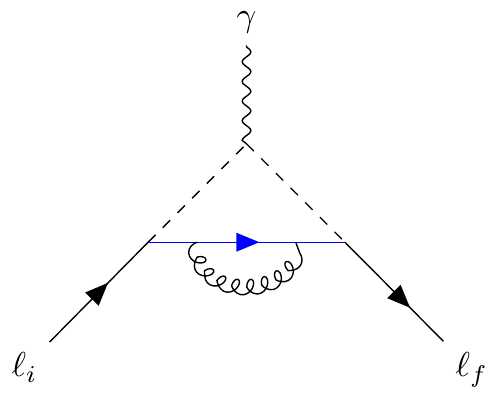}
	\caption{Diagrams sourced at the two-loop level by Eq.~\eqref{yuk}. In total there are 15 different contributions as the topologies which are not specular result in two different contributions. Blue lines indicate a propagating $u$ fermion.  }
	\label{2LQEDLQ}
\end{figure}

In the case of the present analysis, applying the above procedure to the diagrams in Figs.~\ref{1LQEDLQ} and~\ref{2LQEDLQ} reduces the integration of the involved multi-scale Feynman amplitudes to products of single-scale tadpole diagrams. The latter are then easily addressed for instance via the {\tt MATAD}~\cite{Steinhauser:2000ry} package for the {\tt FORM} \cite{Ruijl:2017dtg} symbolic manipulation system, up to the three-loop level in $d$ dimensions. The large-mass expansion is performed by means of the {\tt EXP} and {\tt q2e} codes~\cite{Seidensticker:1999bb,Harlander:1997zb}, prompting the resulting output to {\tt MATAD}. Other technicalities, as the reduction of scalar products of loop momenta belonging to different sub-diagrams or external momenta, are addressed by {\tt FORM} codes developed in-house that implement the integration by parts identities generated through {\tt LiteRed}~\cite{Lee:2013mka}. We use {\tt QGRAF}~\cite{Nogueira:1991ex} to produce the involved diagrams. All computations are performed in the $\rm \overline{MS}$ renormalization scheme.

Before considering the first of the proposed scenarios, we want to remark on the interplay between gauge invariance and ultra-violet divergences that renormalization establishes. Within a renormalizable theory, the available counterterms needed for the regularization of the loop corrections are in 1 to 1 correspondence with the four-dimensional operators contained in the Lagrangian. The renormalizability of the theory specified in Eq.~\eqref{yuk} then ensures that form factors other than $F_1(t)$ are necessarily finite quantities. 

This is certainly the case at the one-loop level, where the relevant corrections are finite. On general grounds, however, we expect that terms proportional to $1/(d-4)$ appear at the two-loop level due to the presence of divergent sub-diagrams. The consistency of the framework requires these terms to cancel upon the inclusion of the diagrams in Fig.~\ref{1LQEDLQ}, properly dressed with renormalized couplings and propagators. In performing our analyses we have explicitly verified these cancellations for a generic choice of the gauge parameter. The expression obtained for each individual amplitude can be found in the ancillary file.  

%-------------------------------------------------------------------------------
\subsection{Scenario I : $M_{u} \gg M_\Phi \gg m_\ell$} \label{sec31}
%---------------------------------------------------------------
We begin our exploration of scalar SLMs with the scenario where the new fermionic degree of freedom $u$ is much heavier than the scalar $\Phi$. The expressions for the analyzed form factors delivered through the detailed computational strategy are 
\begin{equation}
	\label{eq:s1}
	\mathcal{F}_{\rm I} = \chi_{\mathcal{F}}  \frac{m_\ell}{m_{u}} \left\{ \left(N_c \kappa^{\textrm{1L}}_{\rm I} + \alpha_s T_c \kappa^{\textrm{2L}}_{\rm I} \right) +  \frac{m_{\Phi}^2}{m_u^2}\left( N_c \rho^{\textrm{1L}}_{\rm I} +  \alpha_s T_c\rho^{\textrm{2L}}_{\rm I} \right) \right\} ,
\end{equation}
where the specific form factor label $\mathcal{F}_{\rm I}$ and corresponding coefficient $\chi_{\mathcal{F}}$ are given in Tab.~\ref{ts}. In the expression above, $N_c = 3$ is the number of colors in the fundamental representation and $T_c = 4$ is half the trace of the unit matrix in the color adjoint representation. The loop functions are instead specified by  
\bea
\kappa_{\rm{I}}^{\textrm{1L}} &=& - \frac{1 + 2 Q_u}{32 \pi^2}, \\
\rho_{\rm{I}}^{\textrm{1L}} &=& \frac{1}{32 \pi^2} \bigg[ - 3 - 2 Q_u + 4 (1 + Q_u)\ln \frac{m_u}{m_{\Phi}} \bigg], \\
\kappa_{\rm{I}}^{\textrm{2L}} &=& - \frac{30 + \pi^2 + 63 Q_u}{576 \pi^3}, \\
\rho_{\rm{I}}^{\textrm{2L}} &=& \frac{1}{16 \pi^3}\bigg[ -\frac{141 + 2 \pi^2 + 90 Q_u}{72} + 5 (1 + Q_u) \ln ^2 \frac{m_u}{m_\Phi} - 3 \frac{5 + 4 Q_u}{4} \ln \frac{m_\Phi}{\mu}  + \nn \\ &+& \left(- \frac{3}{2} + 3 (1 +  Q_u) \ln \frac{m_\Phi}{\mu} \right) \ln \frac{m_u}{m_\Phi} \bigg],
\eea
which hold in the minimal subtraction scheme.

We observe that the \emph{leading} one-loop contribution\footnote{Other terms at the same loop order have a relative suppression of $(m_\Phi/ m_{u})^2$.} vanishes if $Q_u = -1/2$. 

\begin{table}[htb!]
	\centering
	\begin{tabular}{c|c|c|c|c}
		\toprule
		$\mathcal{F}_{j}:$ & $F_2(0)$  & $G_2(0)$&  $\tilde F_2(0)$  & $\tilde G_2(0)$\\
		\midrule
		$\chi_{\mathcal{F}}:$ 
		&$y_{L_i} y_{R_i}^{\dagger} + y_{L_i}^{\dagger} y_{R_i}$ 
		&$y_{L_i} y_{R_i}^{\dagger} - y_{L_i}^{\dagger} y_{R_i}$ 
		&$y_{L_i} y_{R_f}^{\dagger} + y_{L_f}^{\dagger} y_{R_i} $ 
		&$y_{L_i} y_{R_f}^{\dagger} - y_{L_f}^{\dagger} y_{R_i}$\\[3ex]
		\bottomrule
	\end{tabular}
	\caption{Form factors and corresponding coefficients to be used in Eqs.~\eqref{eq:s1},~\eqref{eq:s2} and~\eqref{eq:s3} for $j = {\rm I, II, III }$, respectively. }
	\label{ts}
\end{table}

Notice also that although the corrections that determine g-2 and EDM differ already at the one loop level, the difference is always proportional to the mass of the involved lepton. In the scenarios we analyse, given the considered mass hierarchy, such difference is therefore negligible. 

%-------------------------------------------------------------------------------
\subsection{Scenario II: $M_{u} \sim M_\Phi \gg m_\ell$}
%---------------------------------------------------------------
\label{hier2}
In this regime the only mass ratio available is ${m_\ell}/{m_{u}}={m_\ell}/{m_{\Phi}}$. Given that observations force $m_\ell$ to be many orders of magnitude below the remaining masses, we safely retain only first order terms in the expansions of the form factors:
\begin{equation}
	\label{eq:s2}
	\mathcal{F}_{\rm II} = \chi_{\mathcal{F}} \frac{m_\ell}{m_{\Phi}}\left(N_c \kappa^{\textrm{1L}}_{\rm II} + \alpha_s T_c \kappa^{\textrm{2L}}_{\rm II} \right) ,
\end{equation}
where $\mathcal{F}_{\rm II}$ and $\chi_{\mathcal{F}}$ are presented in Tab.~\ref{ts}. For this scenario we find 
\bea
\kappa_{\rm{II}}^{\textrm{1L}} &=& \frac{1 + 3 Q_u}{96 \pi^2}, \\
\kappa_{\rm{II}}^{\textrm{2L}} &=& - \frac{1 + 2 Q_u}{384 \pi^3} \left[5 + 3 \ln \frac{m_{\Phi}}{\mu} \right].
\eea
We notice that the one-loop contribution identically vanishes if $Q_u = -1/3$, barring negligible corrections proportional to further powers of the $m_\ell/m_u$ ratio.

%-------------------------------------------------------------------------------
\subsection{Scenario III: $M_{\Phi} \gg M_u \gg m_\ell$}
%---------------------------------------------------------------
Finally, referring to Tab.~\ref{ts}, for the case $M_{\Phi} > M_u$ we have
\begin{equation}
	\label{eq:s3}
	\mathcal{F}_{\rm{III}} = \chi_{\mathcal{F}} \frac{m_\ell}{m_{\Phi}} \left\{ \frac{m_u}{m_{\Phi}}\left(N_c \kappa^{\textrm{1L}}_{\rm{III}} + \alpha_s T_c \kappa^{\textrm{2L}}_{\rm{III}} \right) +  \frac{m_u^3}{m_{\Phi}^3}\left( N_c \rho^{\textrm{1L}}_{\rm{III}} +  \alpha_s T_c\rho^{\textrm{2L}}_{\rm{III}} \right) \right\},
\end{equation}
where the involved loop functions here are:  
\bea
\kappa_{\rm{III}}^{\textrm{1L}} &=& \frac{-1 + 2 Q_u}{32 \pi^2} \left[1 + 2 \ln \frac{m_u}{m_{\Phi}} \right],  \\
\rho_{\rm{III}}^{\textrm{1L}} &=&  \frac{1}{32 \pi^2} \left[-3 + 2 Q_u + 4(2 Q_u - 1) \ln \frac{m_u}{m_{\Phi}}\right],\\
\kappa_{\rm{III}}^{\textrm{2L}} &=& \frac{1}{16 \pi^3}\bigg[ -\frac{87 + 2 \pi^2 - 162 Q_u}{72} + 2 Q_u \ln^2 \frac{m_\Phi}{m_u}   + 3 \frac{1 - 4 Q_u}{4} \ln \frac{m_u}{\mu} + \nn \\ &+&  \left( \frac{3 - 16 Q_u}{4} + 3 Q_u \ln \frac{m_u}{\mu}  \right) \ln \frac{m_\Phi}{m_u} \bigg], \\
\rho_{\rm{III}}^{\textrm{2L}} &=& \frac{1}{8 \pi^3}\bigg[-\frac{138 + \pi^2 - 99 Q_u}{72} + 2 Q_u \ln ^2 \frac{m_\Phi}{m_u}  + 3 (1 -  Q_u) \ln \frac{m_u}{\mu}   + \nn \\ &+&  \left(\frac{3}{2} (2 - 3 Q_u) - 3  (1 - 2 Q_u) \ln \frac{m_u}{\mu} \right)\ln \frac{m_\Phi}{m_u} \bigg] .
\eea

We remark that scalar leptoquarks models fall in this scenario after the $u$ fermion is identified, upon a Majorana conjugation, with the SM top quark. 

%-------------------------------------------------------------------------------
\section{Precision tests of simplified leptoquark models}\label{plots}
%-------------------------------------------------------------------------------

The form factors in the previous sections provide a straightforward way to match the lepton precision observables within any scheme of new physics that recovers, in a limit, the proposed SLMs. In more detail, possible radiative flavor violating decays can be tested against the known limits by simply computing 
\bea \label{br}
\textrm{Br}(\ell_i \rightarrow \ell_f \gamma) = \frac{\alpha_{em} m_{\ell_i} \tau_{\ell_i}}{2}\left(|\tilde{F}_2(0)|^2 + |\tilde{G}_2(0)|^2 \right),
\eea
where $\alpha_{em}$ is the fine structure constant and  $m_{\ell_i}$, $\tau_{\ell_i}$ the mass and lifetime of the initial state lepton, respectively.
The observable plays an important role in scenarios where new degrees of freedom couple to different SM generations and strongly constrains, for instance, any simultaneous explanation of the electron and muon AMM anomalies.  The inclusion of new complex Yukawa couplings in Eq.~\eqref{yuk} furthermore provides additional sources of CP-violations, which inevitably contribute to the EDM $d_{\ell} = i\, G_2(0)/2 m_\ell $. Together with the AMM $a_\ell = F_2(0)$, the mentioned observables offer a way to fully test the Yukawa sector in Eq.~\eqref{yuk} and, more in general, any scalar leptoquark sector of new physics models.
 
Pending the release of new data concerning the muon AMM, in this paper we opt to focus on the impact of $(g-2)_{\mu,e}$ to demonstrate the reach of the proposed framework. We pay particular attention to the identification of regions in the parameter space where the two-loop contributions are comparable to the LO ones, because of an enhancement of the former or a suppression of the latter. For the similarity in the loop structures of the analyzed form factors, we expect similar effects to manifest also in the remaining precision observables. 

Because of the tight bound posed by the non-observation of $\mu \rightarrow \gamma e$ transitions, on top of the absence of an efficient suppression mechanism, we a priori disregard common explanations of the $(g-2)_{e,\mu}$ anomalies. Similarly, the tight bound on the $d_e$ forces aligned phases in the Yukawa couplings for the electron in Eq.~\eqref{yuk}. The case of the muon is different as the current limit does not significantly constrain the corresponding CP phase and, in principle, generic Yukawa parameters are thus allowed. In regard of this, the future increment of 2 order of magnitudes in the sensitivities of the Fermilab and J-PARC experiments~\cite{Semertzidis:2001,PhysRevLett.93.052001}, or the at-least 3 order improvement proposed with PSI~\cite{Adelmann:2010}, will exhaustively probe the possible correlations between large values of $d_{\mu}$ and the muon AMM anomaly that complex Yukawa couplings source~\cite{Crivellin:2018qmi}.    

Before detailing our results, we stress that the performed numerical computations account for the running of SM couplings, but neglect the renormalization group (RG) evolution of the new physics interactions. The residual dependence of form factors on the RG scale $\mu$ is addressed by setting the latter to the value of the intermediate mass in the scenario under consideration. We have checked that the implied uncertainty is negligible for the span of the hierarchy in the masses of the involved colored particles. 

\subsection{Scenario I: $m_u \gg m_{\Phi}\gg m_\ell$}

\begin{figure}[h]
	\centering
	\includegraphics[width=.35\linewidth]{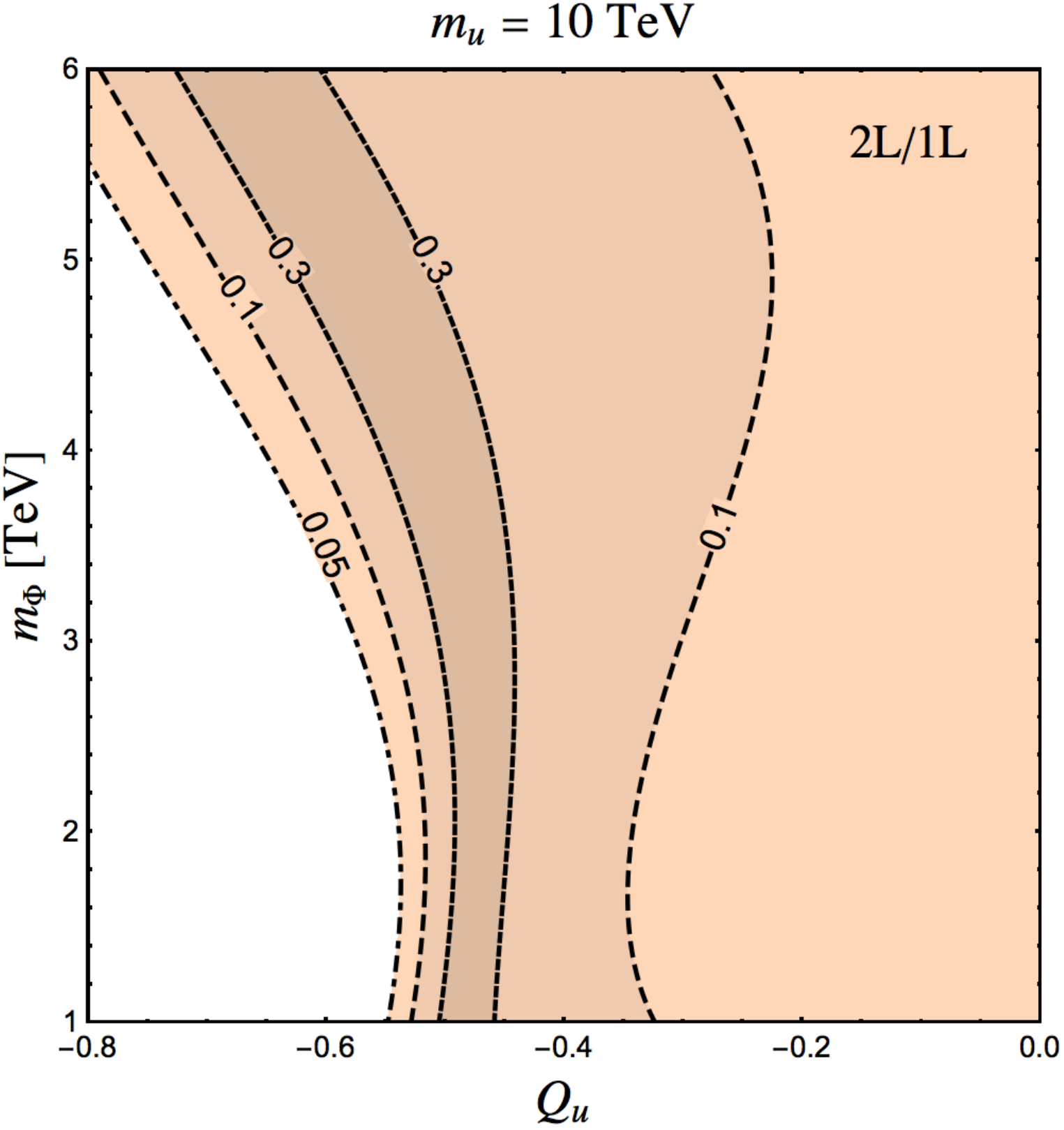}  
	\hspace{1cm}
	\includegraphics[width=.365\linewidth]{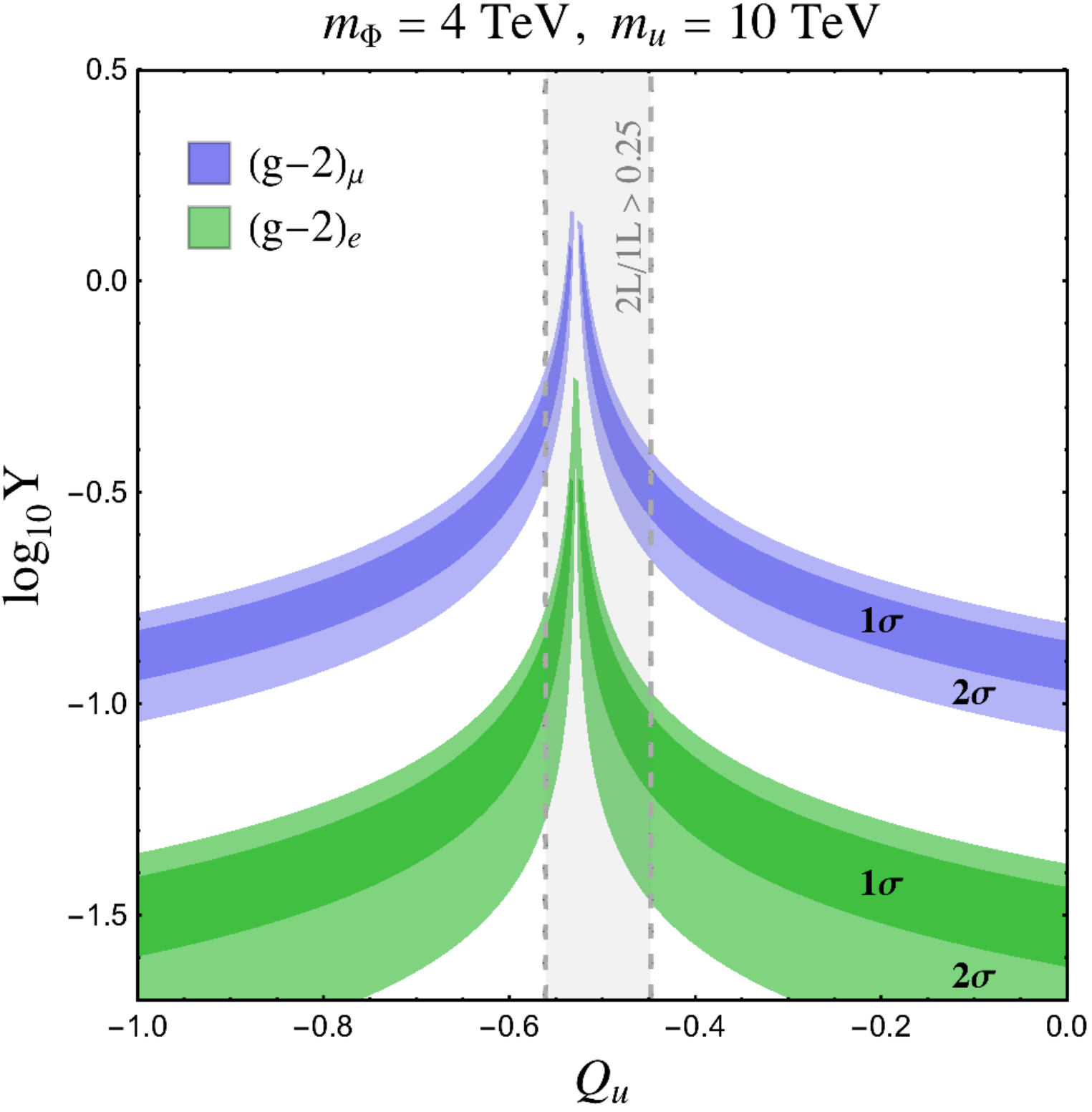}
		\caption{Left panel: magnitude of the two-loop correction relative to the one-loop leading order, as a funtion of the scalar mass $m_\phi$ and the $u$ fermion electric charge $Q_u$. Right panel:  1$\sigma$ and 2$\sigma$ confidence regions for the $(g-2)$ anomaly of electron (green) and muon (blue) evaluated at the NLO as a function of the Yukawa coupling, $Y = \sqrt{|Y_L| |Y_R|}$, and $Q_u$, assuming $m_u = 10$ TeV and $m_{\Phi} = 4$ TeV. The gray band centered on $Q_u = -1/2$ signals the region where the two-loop contribution cannot be neglected.}
	\label{LQ1plots1}
\end{figure}

In the first panel of Fig.~\ref{LQ1plots1} we show the magnitude of the two-loop contribution relative to the one-loop correction, as a function of the intermediate mass $m_\Phi$ of the scalar field and the charge $Q_u$ of the heavy fermion $u$. Because the leading contribution into the one-loop result vanishes for $Q_u = -1/2$, we observe the presence of a region centered around  this critical value where the NLO cannot be neglected. The effect broadens as the ratio $m_{u}/m_{\Phi}$ is relaxed and is therefore of interest for the phenomenology of UV-completed models that contain extra fermions on top of scalar leptoquarks~\cite{Hewett:1997ba,Babu:1997tx}.  

\begin{figure}[t]
	\centering
	\includegraphics[width=.35\linewidth]{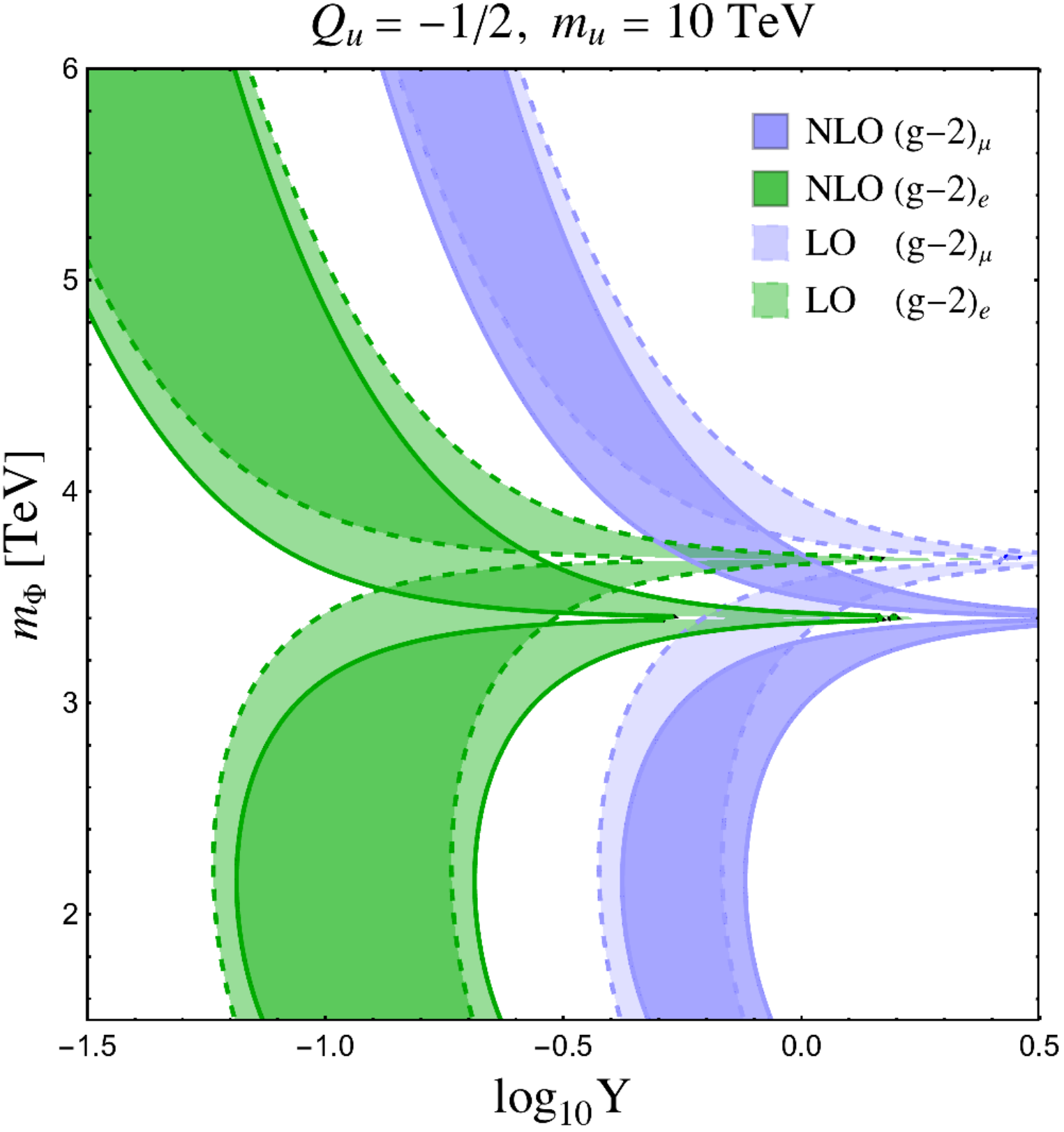}
	\hspace{1cm}
	\includegraphics[width=.35\linewidth]{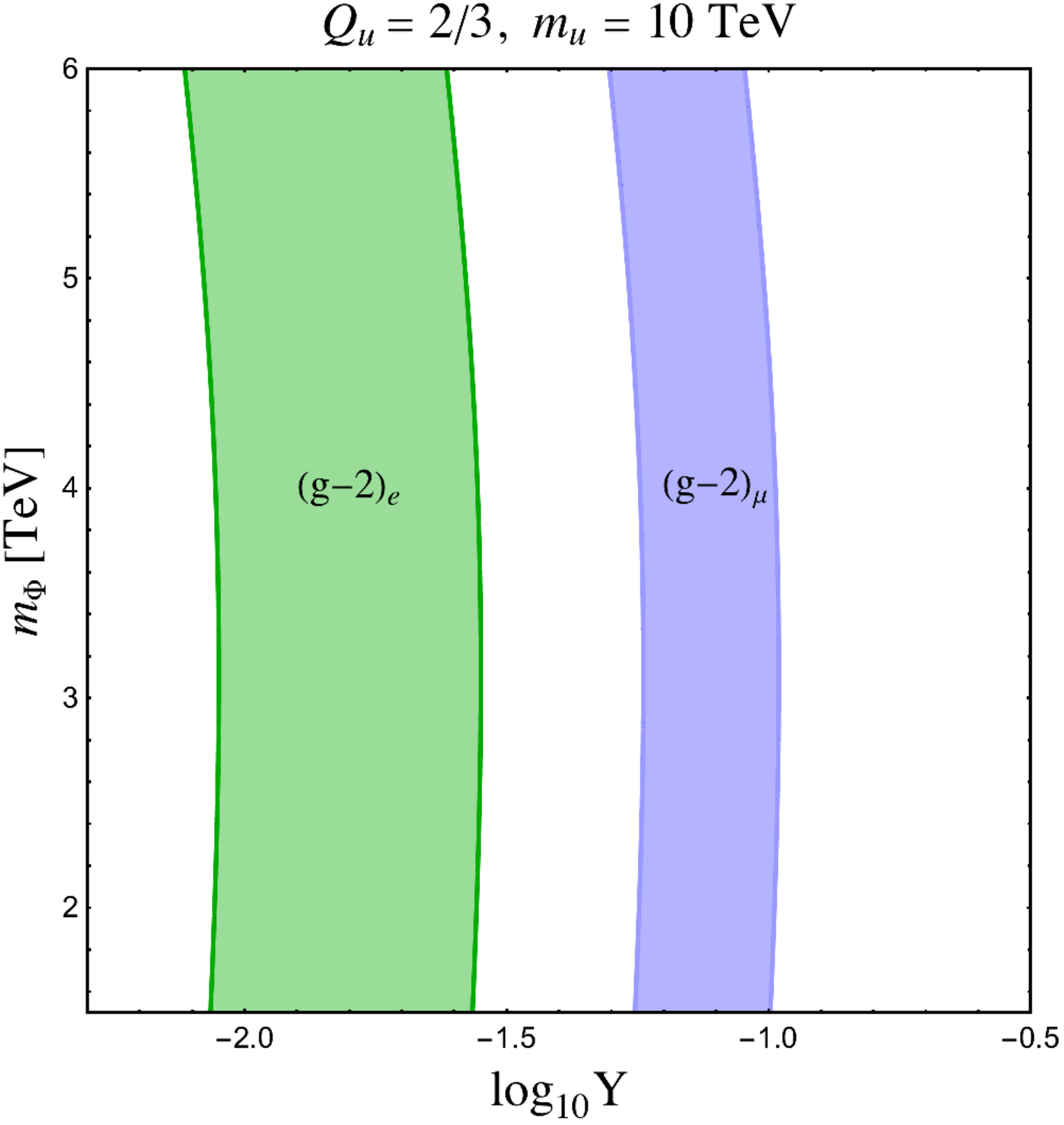}
		\caption{Left panel: $2\sigma$ contours for the AMM of electron (green) and muon (blue), as a function of the new lepton Yukawa coupling $Y = \sqrt{|Y_L| |Y_R|}$ and the mass of the involved scalar field $m_\phi$. The charge of the heavy fermion is set at the critical value $Q_u = - 1/2$. The lighter regions correspond to the prediction of the LO contribution only, whereas the darker region account also for the NLO result. Right panel: same as the left panel for a non-critical charge $Q_u=2/3$. The solutions for NLO are indistinguishable from those of the LO alone.}  
	\label{LQ1plots2}
\end{figure}

\begin{figure}[h]
	\centering
	\includegraphics[width=.35\linewidth]{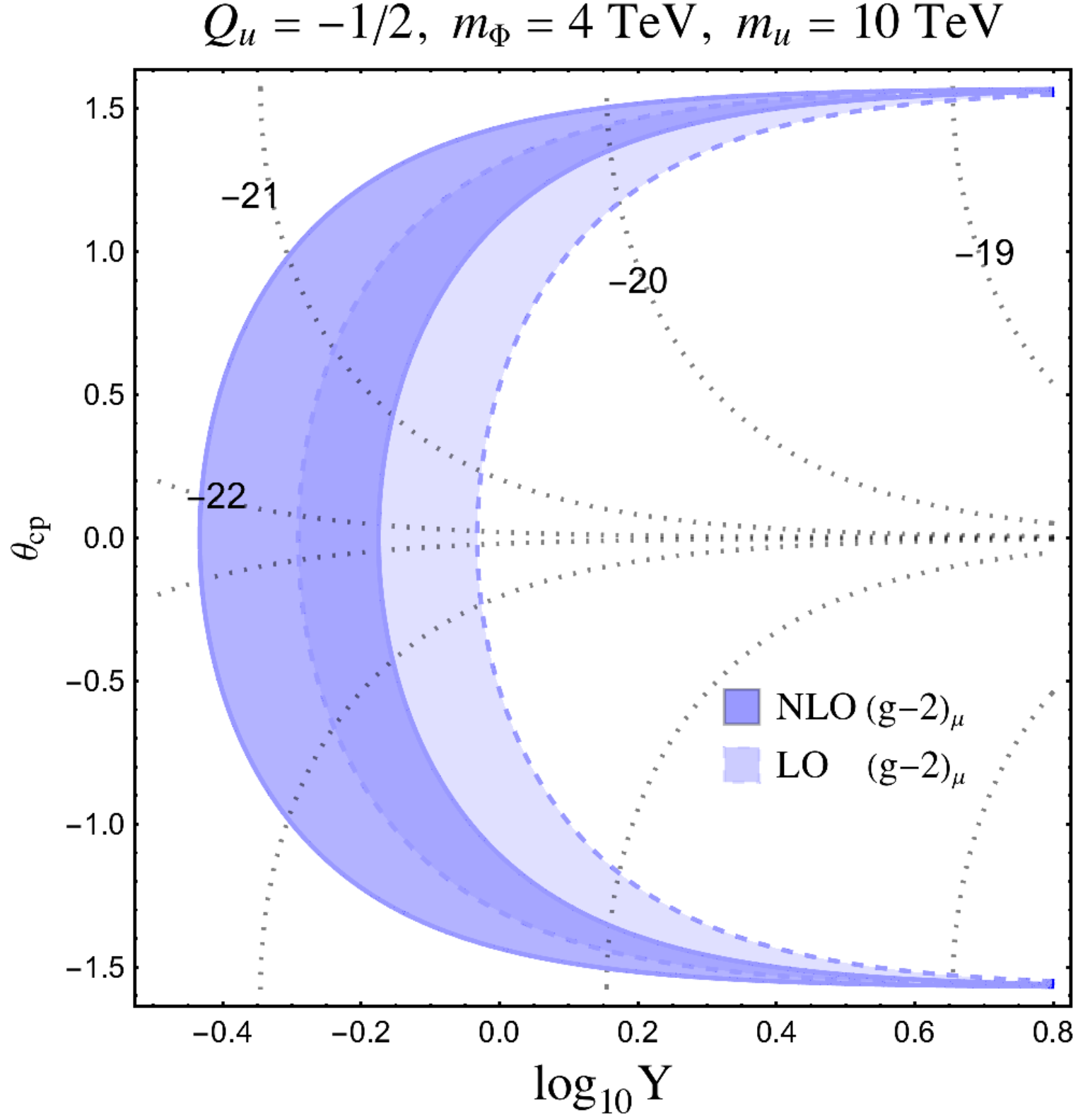}
	\hspace{1cm}
	\includegraphics[width=.35\linewidth]{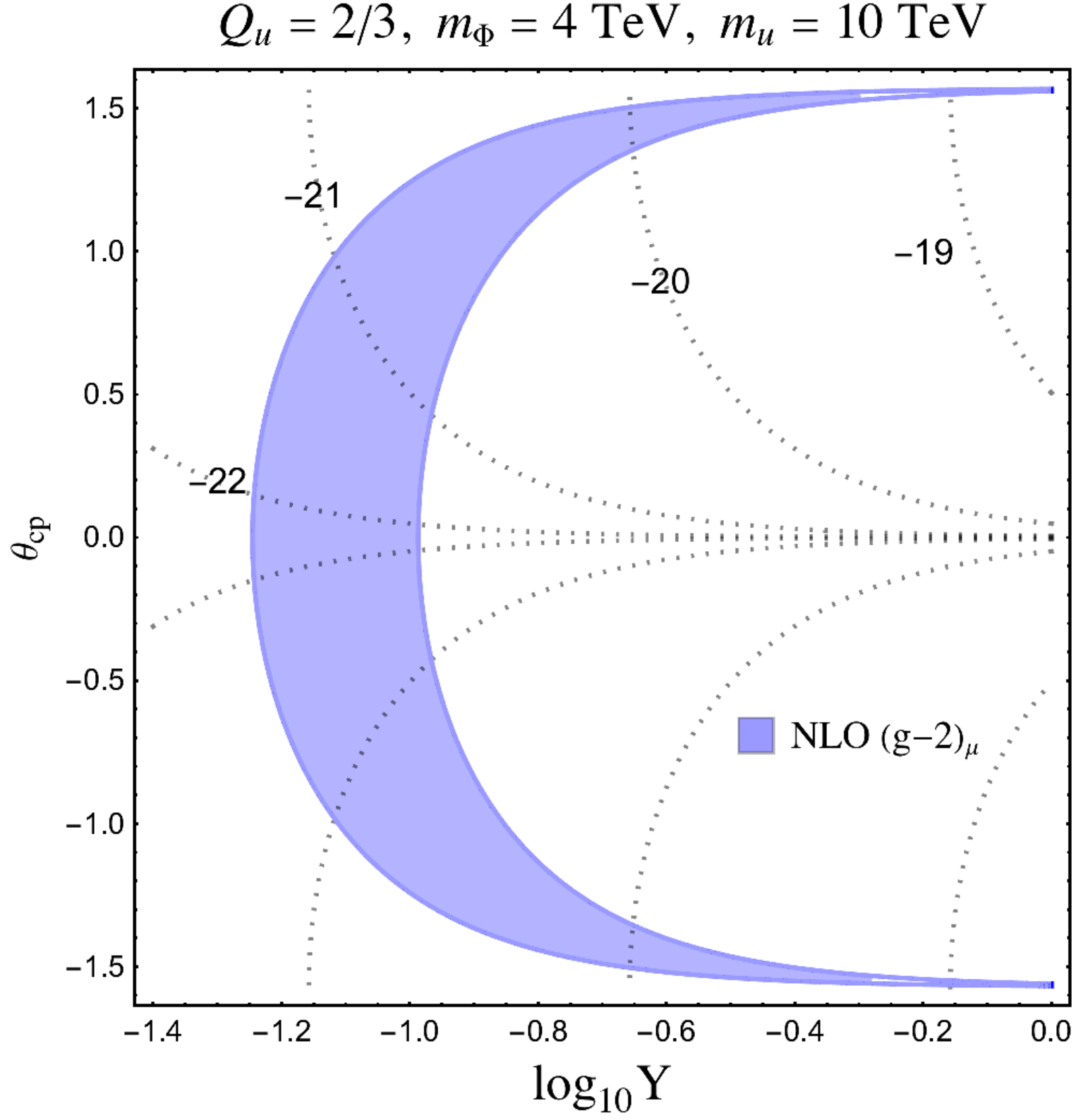}
		\caption{Phase and magnitude of the complex Yukawa coupling of the muon that match, in correspondence of the shaded regions, the measured $(g-2)_{\mu}$ $2\sigma$ interval. In both the panels, the dotted lines indicate the expected order of magnitude of the muon EDM ($e\cdot$cm). Left panel: $Q_u = -1/2$, the LO solutions (lighter area) are clearly distinguishable from the full two-loop contribution (darker). Right panel: $Q_u = 2/3$, the NLO solutions are indistinguishable from the LO ones.}
	\label{LQ1plots3}
\end{figure}
\FloatBarrier

The mentioned region is plotted again in the right panel of Fig.~\ref{LQ1plots1}, where a gray vertical band indicates the region of parameter space in which the two-loop term exceeds 1/4 of the corresponding one-loop contribution. The relative size of the considered corrections depends only on the charge $Q_u$ as the Yukawa structure is common to both the terms. The threshold has been chosen arbitrarily to highlight the impact of NLO effects. The dark (light) green and blue areas show the 1$\sigma$ ($2\sigma$) confidence intervals for the electron and muon AMM measurements, respectively. The contours are plotted with respect to the charge of the heavy fermion, $Q_u$, and the lepton-$\Phi$ Yukawa coupling $Y = \sqrt{|Y_L| |Y_R|}$. For the purpose of the plot we assumed a vanishing relative phase. We remark that the bound due to $\mu\to e \gamma$ searches prevents a common explanation of the two anomalies within the present framework.

The impact of the NLO contribution on the AMM of muon and electron is analyzed once again in Fig.~\ref{LQ1plots2}. In the left panel we match the 2$\sigma$ contours for the electron (green) and muon (blue) measurements after setting the $u$ charge to the critical value $Q_u=-1/2$. The lighter areas, which account for the contribution of the LO only, differ significantly from the full NLO solutions indicated by the darker regions. In the right panel we repeat the exercise for a value of the charge $Q_u$ away from the critical one, finding that the LO solutions are essentially indistinguishable from the NLO ones.   

As mentioned before, the current bound on the muon EDM, Eq.~\eqref{dmulim}, leaves space for new complex Yukawa couplings that would induce a correlation between the muon EDM and AMM. In fact, the definitions of the involved form factors show that for complex couplings 
\begin{align}
	a_{\mu} \sim Re\, Y_L Y_R^{\dagger} \sim |Y_L||Y_R|\cos (\theta_L - \theta_R); \qquad
	d_{\mu} \sim Im\, Y_L Y_R^{\dagger} \sim |Y_L||Y_R|\sin (\theta_L - \theta_R).
\end{align}
By opportunely redefining the involved phases we then obtain
\begin{align}
	a_{\mu} \sim Y^{2} \cos (\theta_{cp}); \qquad 	d_{\mu} \sim Y^{2} \sin (\theta_{cp}),
\end{align}
where again  $Y = \sqrt{|Y_L| |Y_R|}$.
In Fig.~\ref{LQ1plots3} we plot these quantities as a function of the magnitude and relative phase of the involved Yukawa couplings, for a critical (left panel) and non-critical (right panel) $Q_u$ charge. In the former case it is clearly possible to distinguish the effect of the two-loop contribution, which could be therefore probed in the forthcoming experiments that target the muon EDM~\cite{Semertzidis:2001,PhysRevLett.93.052001,Adelmann:2010}.

\subsection{Scenario II: $m_{\Phi} \sim m_u\gg m_\ell$}

When the masses of the new colored states are close in value, the  phenomenology of SLMs depends on the unique new physics scale $m_{\Phi} = m_{u}$. Collider experiments force a strong suppression of the ratio $m_\ell/m_{\Phi}$, in a way that the LO always dominates the mass expansion of the considered amplitudes and the structure of the form factors simplifies as shown in Sec.~\ref{hier2}.

\begin{figure}[h]
	\centering
	\includegraphics[width=.35\linewidth]{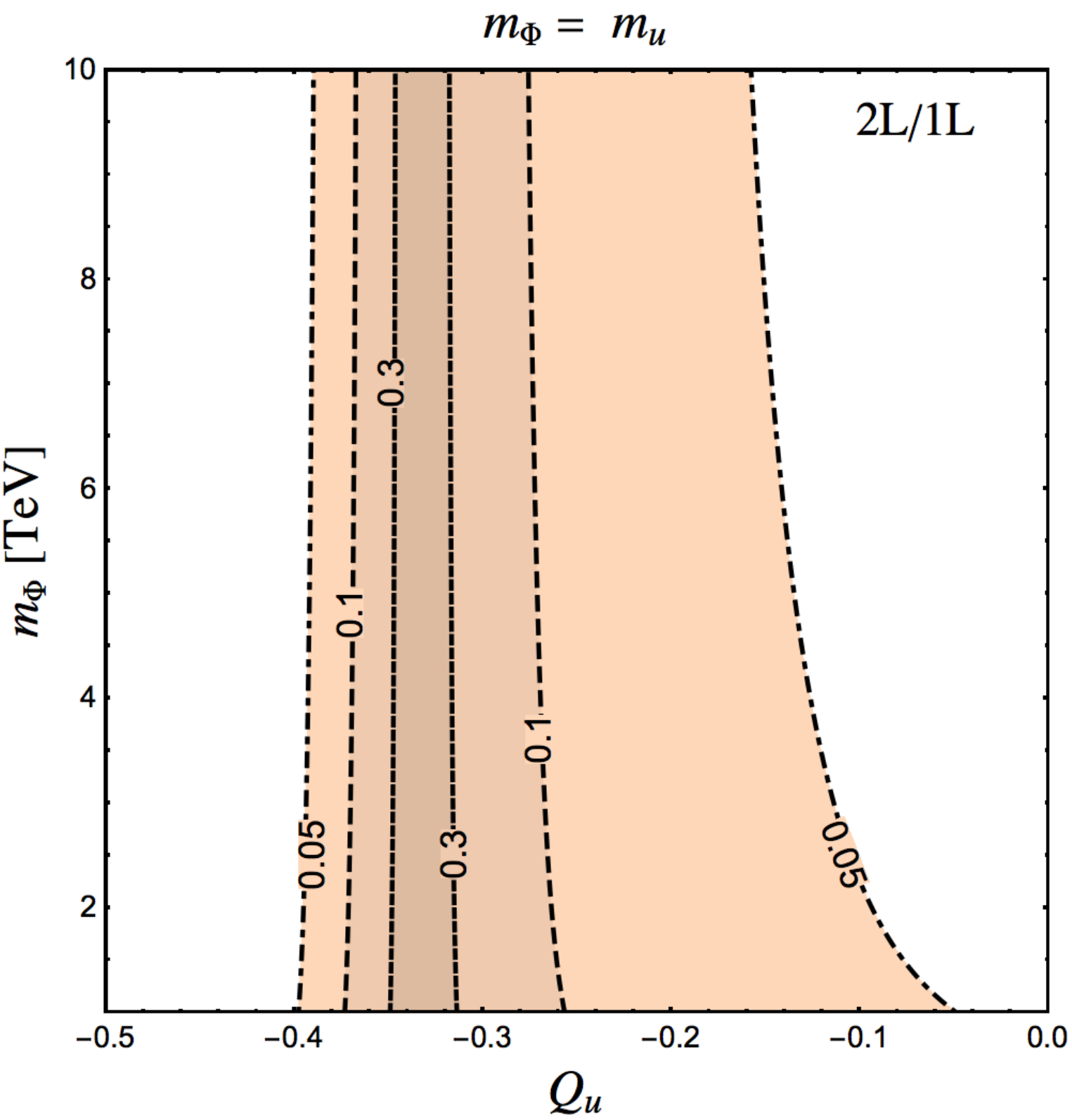}
	\hspace{1cm}
	\includegraphics[width=.36\linewidth]{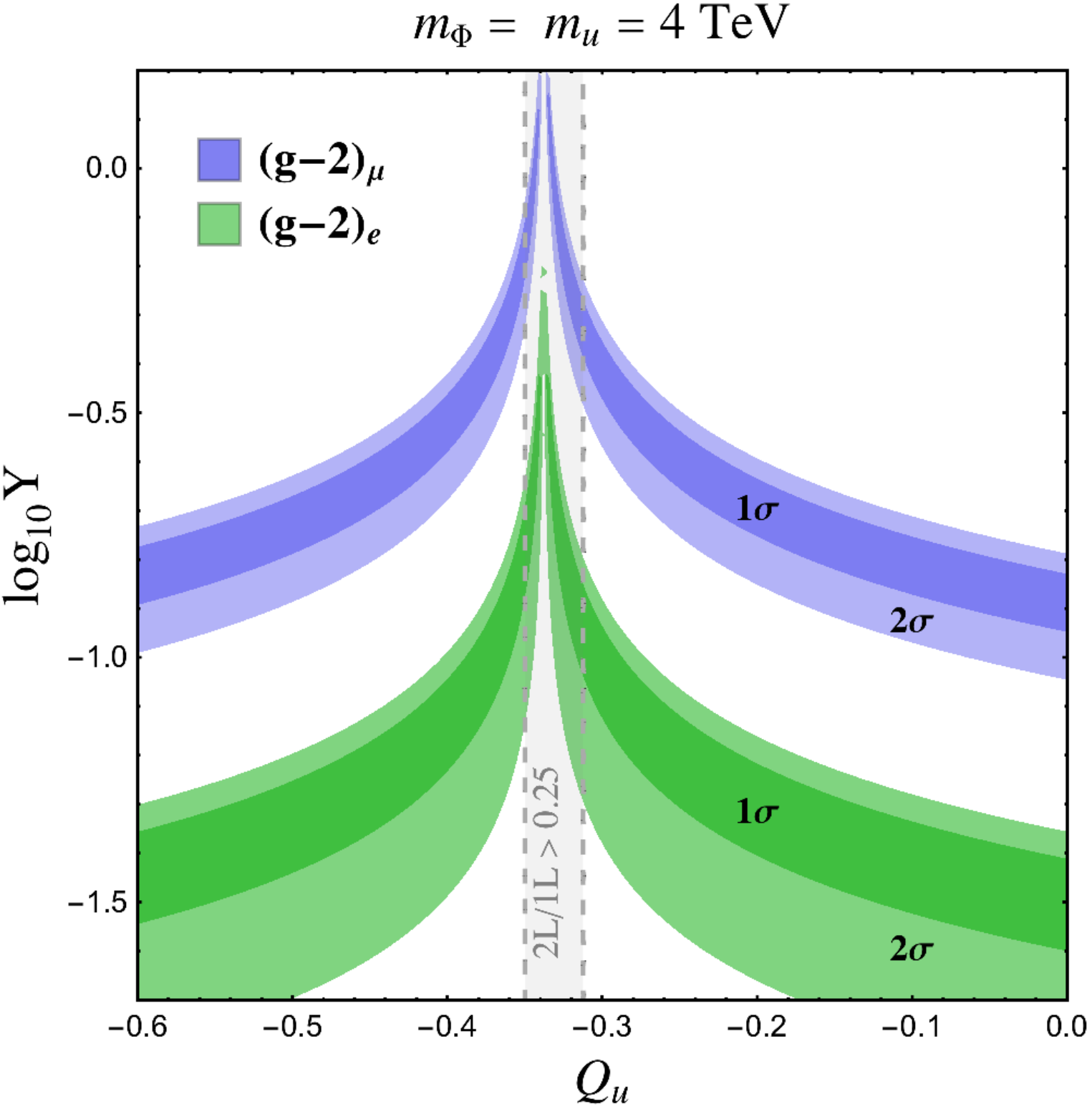}
		\caption{Left panel: magnitude of the two-loop correction relative to the one-loop leading order on the considered parameter space. Right panel:  1$\sigma$ and 2$\sigma$ confidence regions for the $(g-2)$ anomaly of electron (green) and muon (blue) evaluated at the NLO as a function of the Yukawa coupling, $Y$, and electric charge of the $u$ fermion, $Q_u$, assuming $m_u = m_{\Phi} = 4$ TeV. The gray band centered on $Q_u = -1/3$ shows the region where the two-loop contribution cannot be neglected.}
	\label{LQ2plots1}
\end{figure}

In this case, we find that the two-loop contribution is negligible except in a narrow area of the parameter space centered on $Q_u=-1/3$, where the one-loop contribution identically vanishes. This is illustrated in the left panel of Fig.~\ref{LQ2plots1}, where again we plot the magnitude of the two-loop contribution relative to the LO one on the considered parameter space. 

\FloatBarrier

\begin{figure}[h]
	\centering
	\includegraphics[width=.36\linewidth]{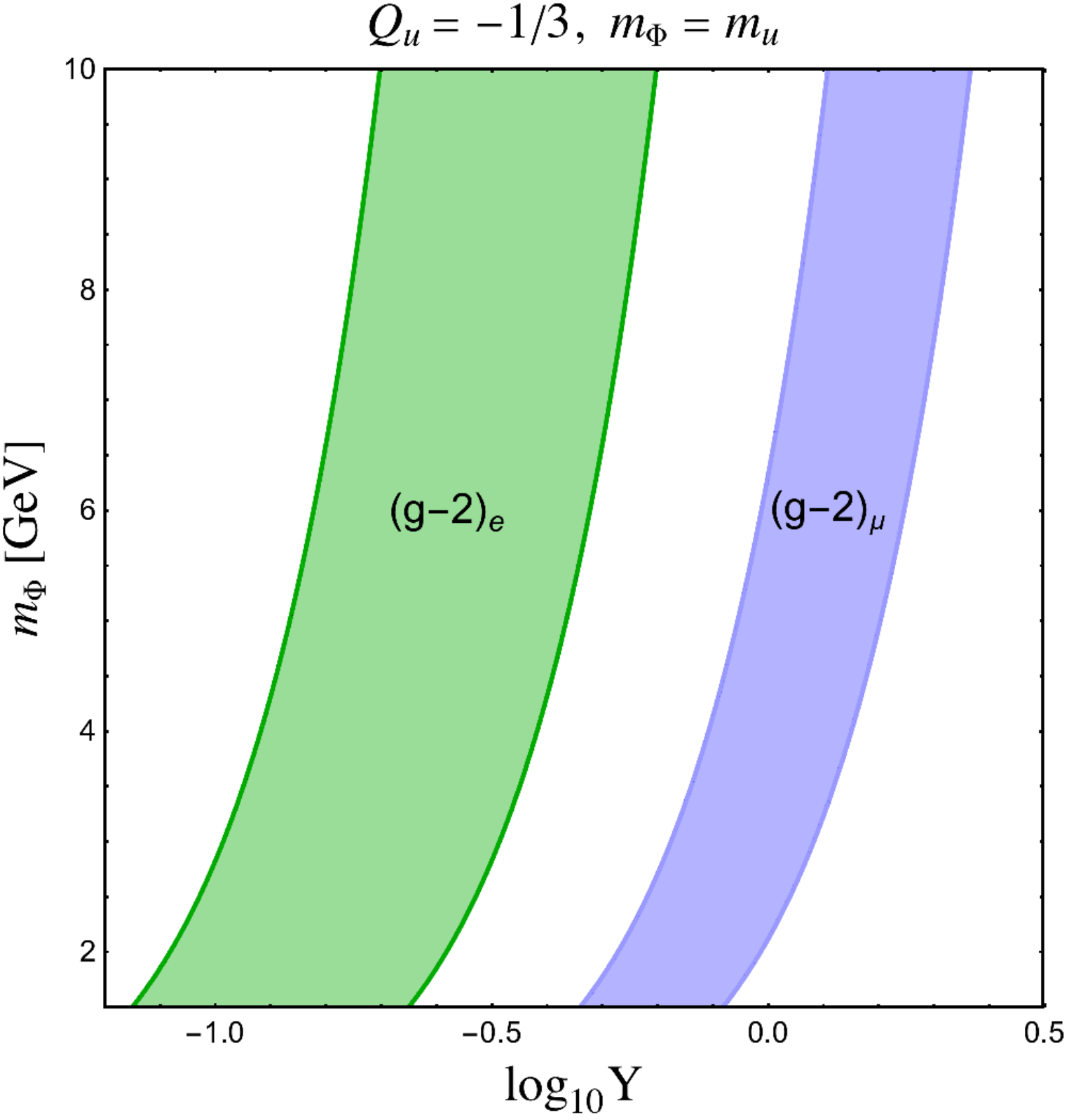}
	\hspace{1cm}
	\includegraphics[width=.35\linewidth]{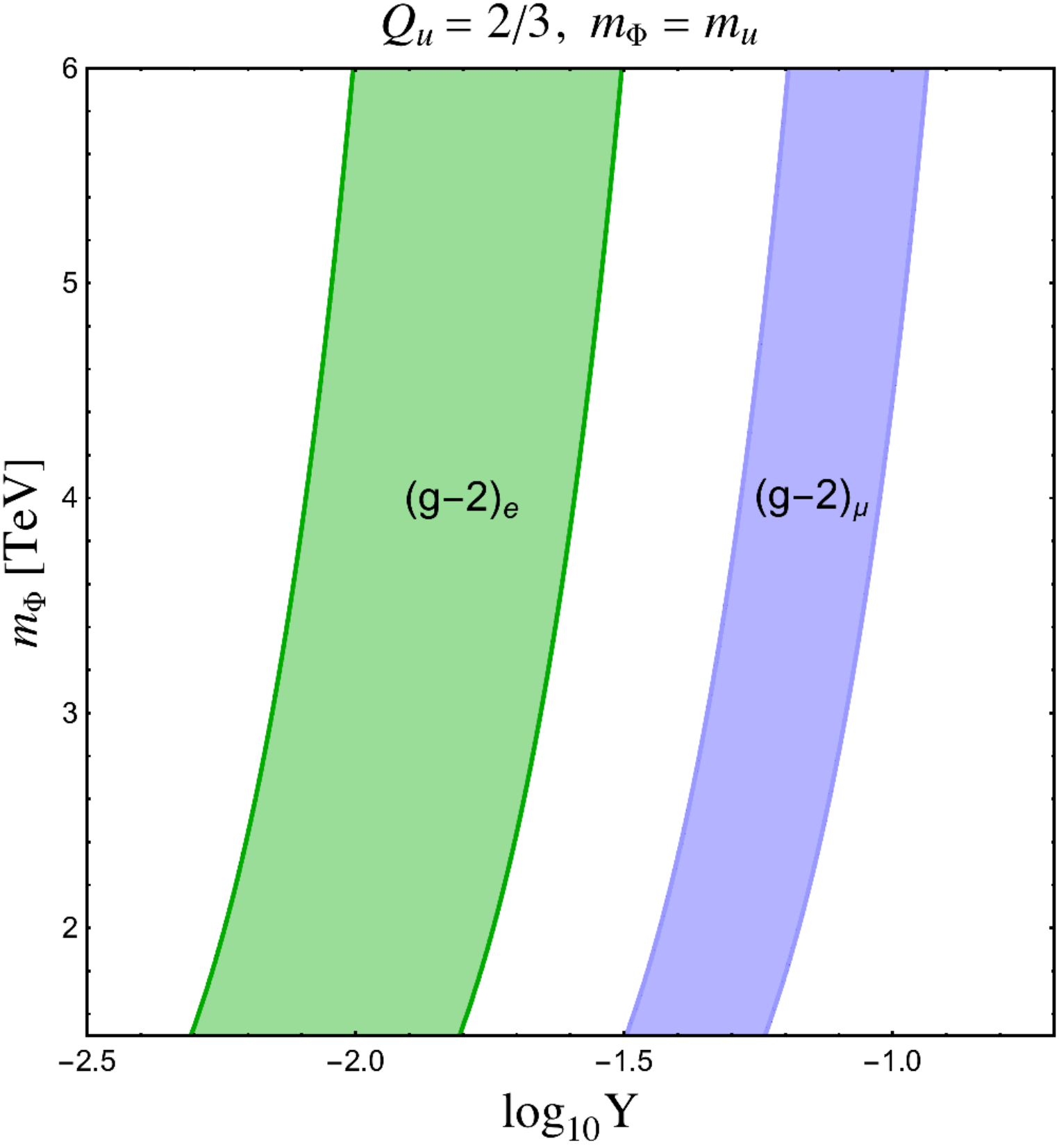}
		\caption{Left panel: $2\sigma$ contours for the AMM of electron (green) and muon (blue), as a function of the lepton Yukawa coupling $Y = \sqrt{|Y_L| |Y_R|}$ and the new physics scale. The charge of the heavy fermion is set at the critical value $Q_u = - 1/3$, forcing the one-loop contribution to identically vanish. Right panel: same as the left panel for a charge $Q_u=2/3$ away form the critical value. The solutions for NLO are indistinguishable from those of the LO alone.}
	\label{LQ2plots2}
\end{figure}

\begin{figure}[h]
\centering
	\includegraphics[width=.35\linewidth]{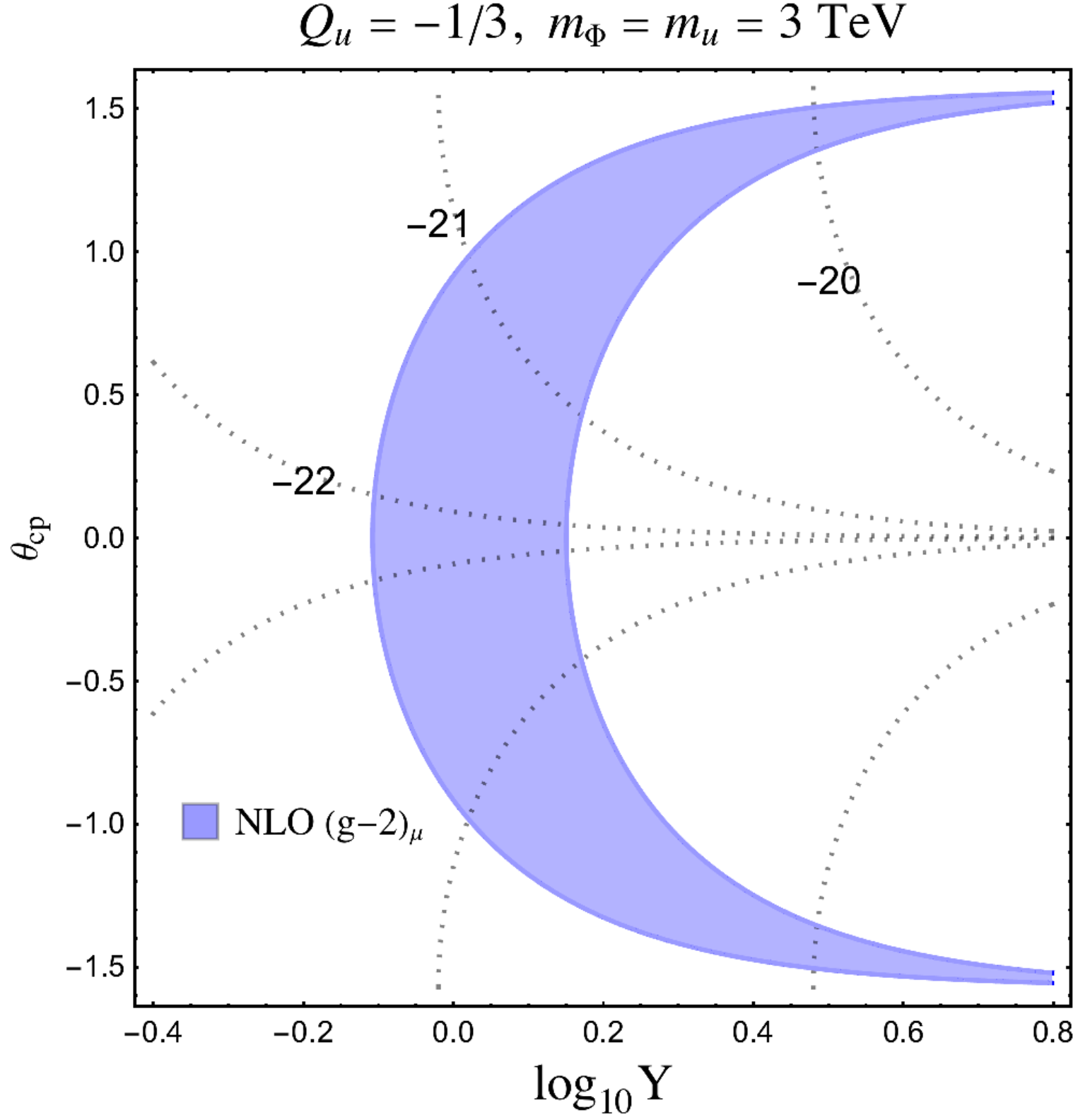}
	\hspace{1cm}
	\includegraphics[width=.35\linewidth]{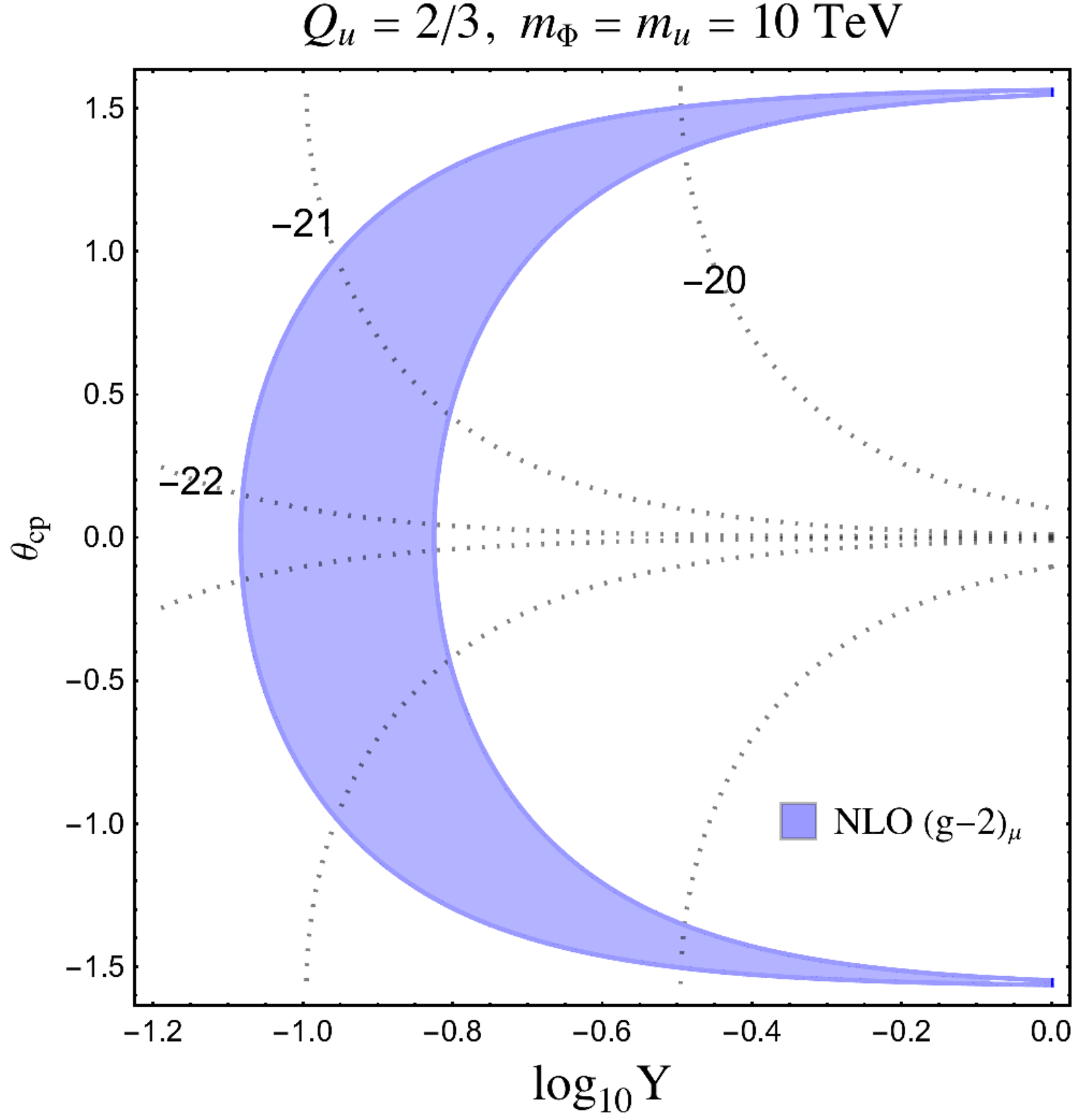}
		\caption{Phase and magnitude of the complex Yukawa coupling of the muon that match, in correspondence of the shaded regions, the measured $(g-2)_{\mu}$ $2\sigma$ interval. In both the panels, the dotted lines indicate the expected order of magnitude of the muon EDM ($e\cdot$cm). Left panel: $Q_u = -1/3$, the solutions are determined by the two-loop contribution alone for a new physics scale of 3 TeV. Right panel: $Q_u = 2/3$ and the new physics scale is set to 10 TeV. The NLO solutions are indistinguishable from the LO ones. }
	\label{LQ2plots3}
\end{figure}

As we can see from the right panel of Fig.~\ref{LQ2plots1} and Figs.~\ref{LQ2plots2},~\ref{LQ2plots3}, the results obtained for the Scenario I hold also in the present case, albeit a different critical value of the charge $Q_u$. We however remark that, at the critical value $Q_u=-1/3$, the precision observables probe here the two-loop contribution alone, rather than a combination of one-loop and two-loop corrections as in the previous scenario.

\FloatBarrier

\subsection{Scenario III: $m_{\Phi} \gg m_u\gg m_\ell$}

\begin{figure}[h]
	\centering
	\includegraphics[width=.35\linewidth]{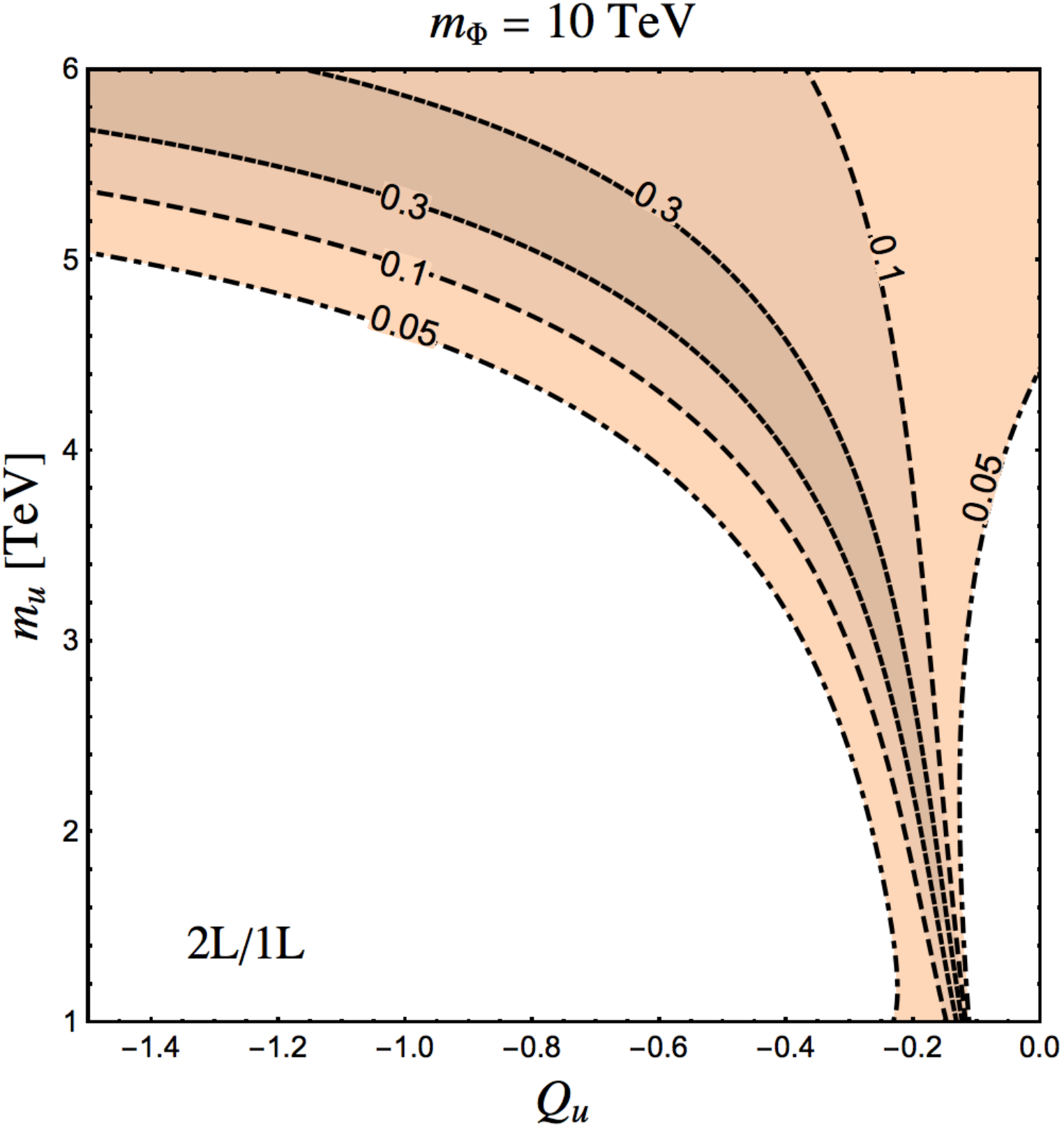}     
	\hspace{1cm} 
	\includegraphics[width=.365\linewidth]{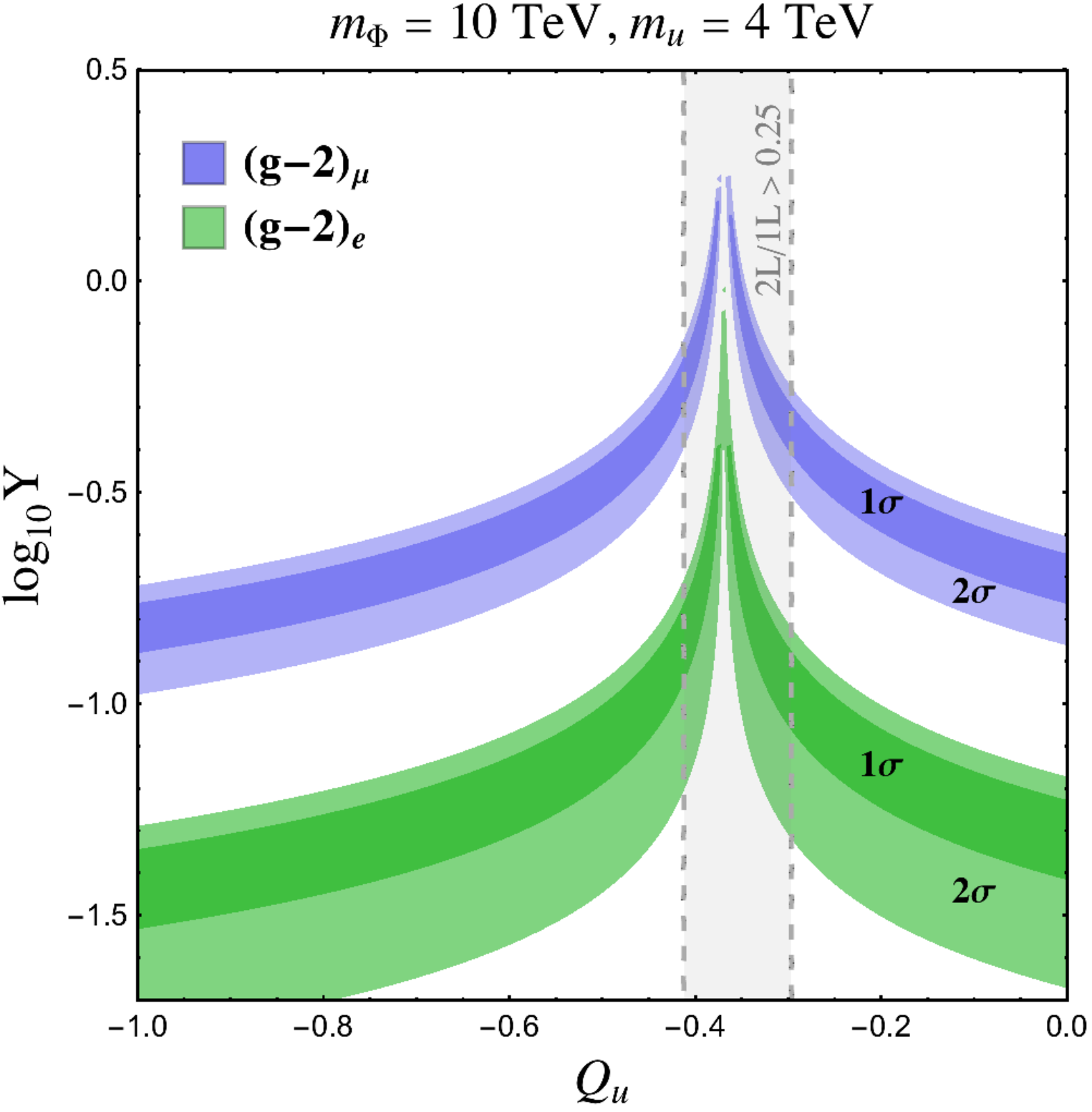}
		\caption{Left panel: magnitude of the two-loop correction relative to the one-loop leading order on the considered parameter space. Right panel:  1$\sigma$ and 2$\sigma$ confidence regions for the $(g-2)$ anomaly of electron (green) and muon (blue) evaluated at the NLO as a function of the Yukawa coupling, $Y = \sqrt{|Y_L| |Y_R|}$, and electric charge of the $u$ fermion, $Q_u$, assuming $m_u = 4$ TeV and $m_\phi = 10$ TeV. The gray band indicates the presence of sizeable NLO contributions.}
	\label{LQ3plots1}
\end{figure}

\begin{figure}[h]
	\centering
	\includegraphics[width=.35\linewidth]{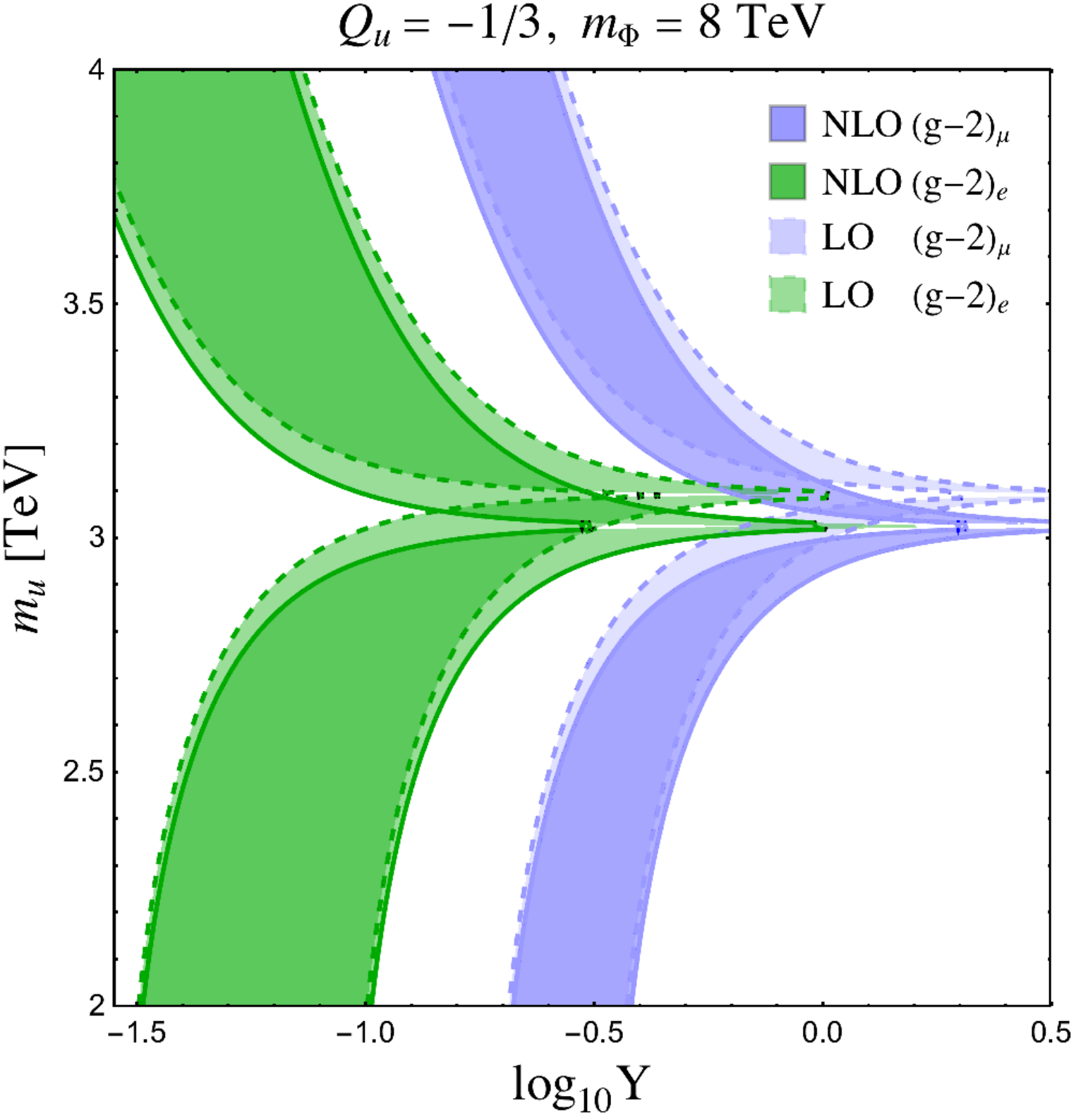}
	\hspace{1cm}
	\includegraphics[width=.35\linewidth]{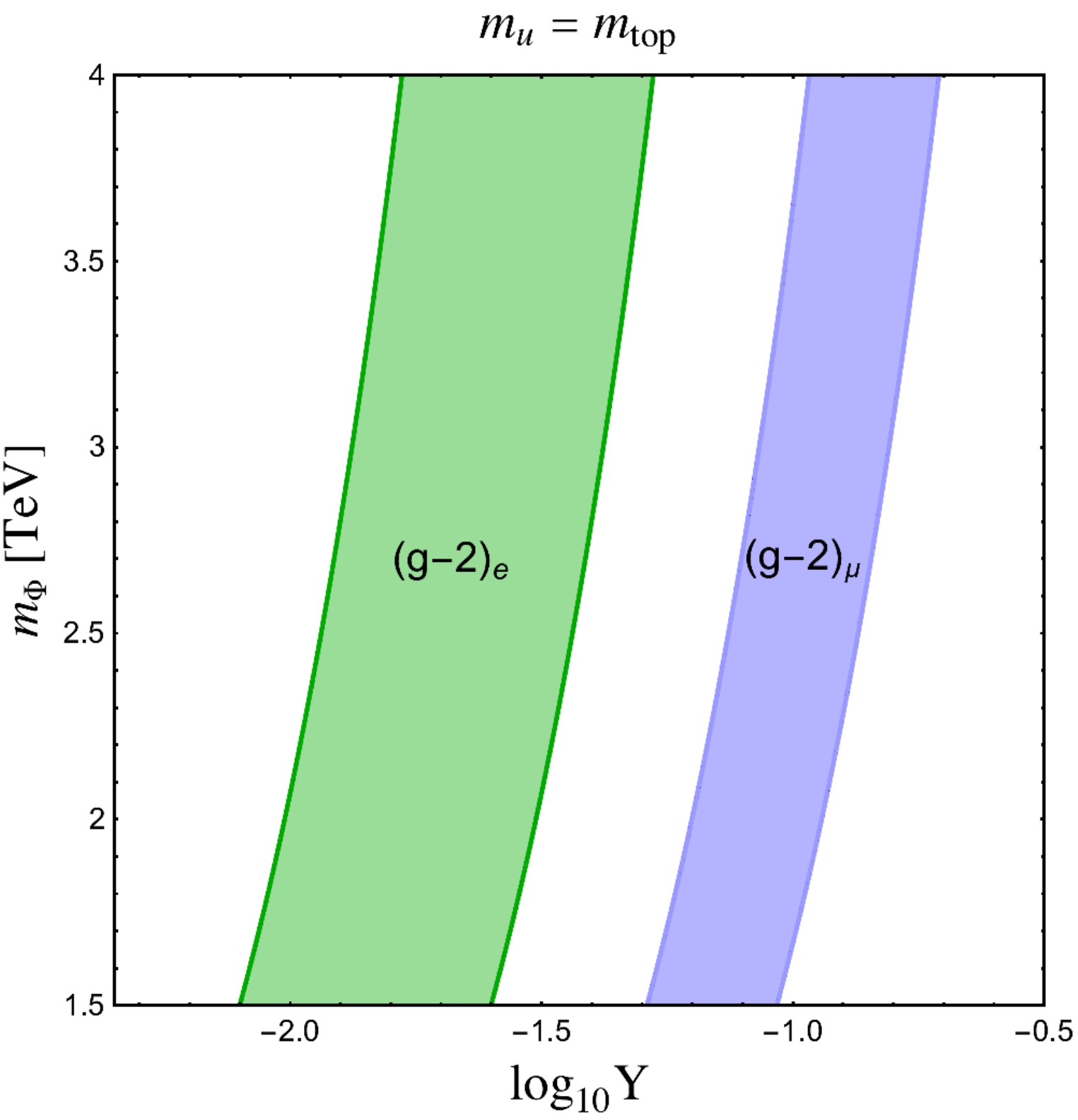}
		\caption{Left panel: $2\sigma$ contours for the AMM of electron (green) and muon (blue), as a function of the lepton Yukawa coupling $Y = \sqrt{|Y_L| |Y_R|}$ and the fermion mass $m_u$ for a critical value of the fermion charge $Q_u = - 1/3$. The NLO solutions are clearly distinguishable from the LO result. Right panel: same as the left panel after identifying the fermion $u$ with the SM top quark. The solutions for NLO are now indistinguishable from those of the LO alone.}
\label{LQ3plots2}
\end{figure}

\begin{figure}[h]
	\centering
	\includegraphics[width=.36\linewidth]{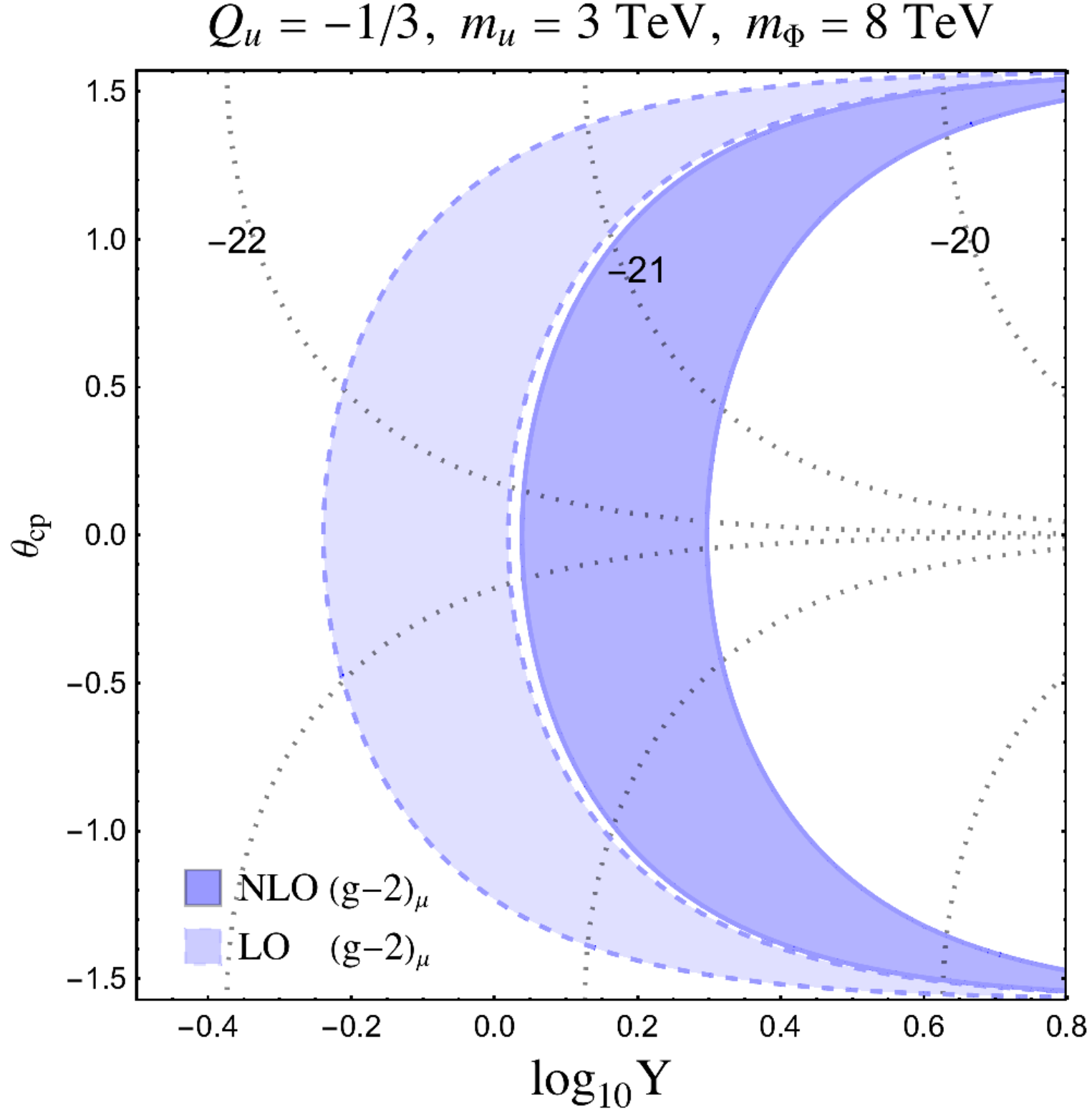}
	\hspace{1cm}
	\includegraphics[width=.35\linewidth]{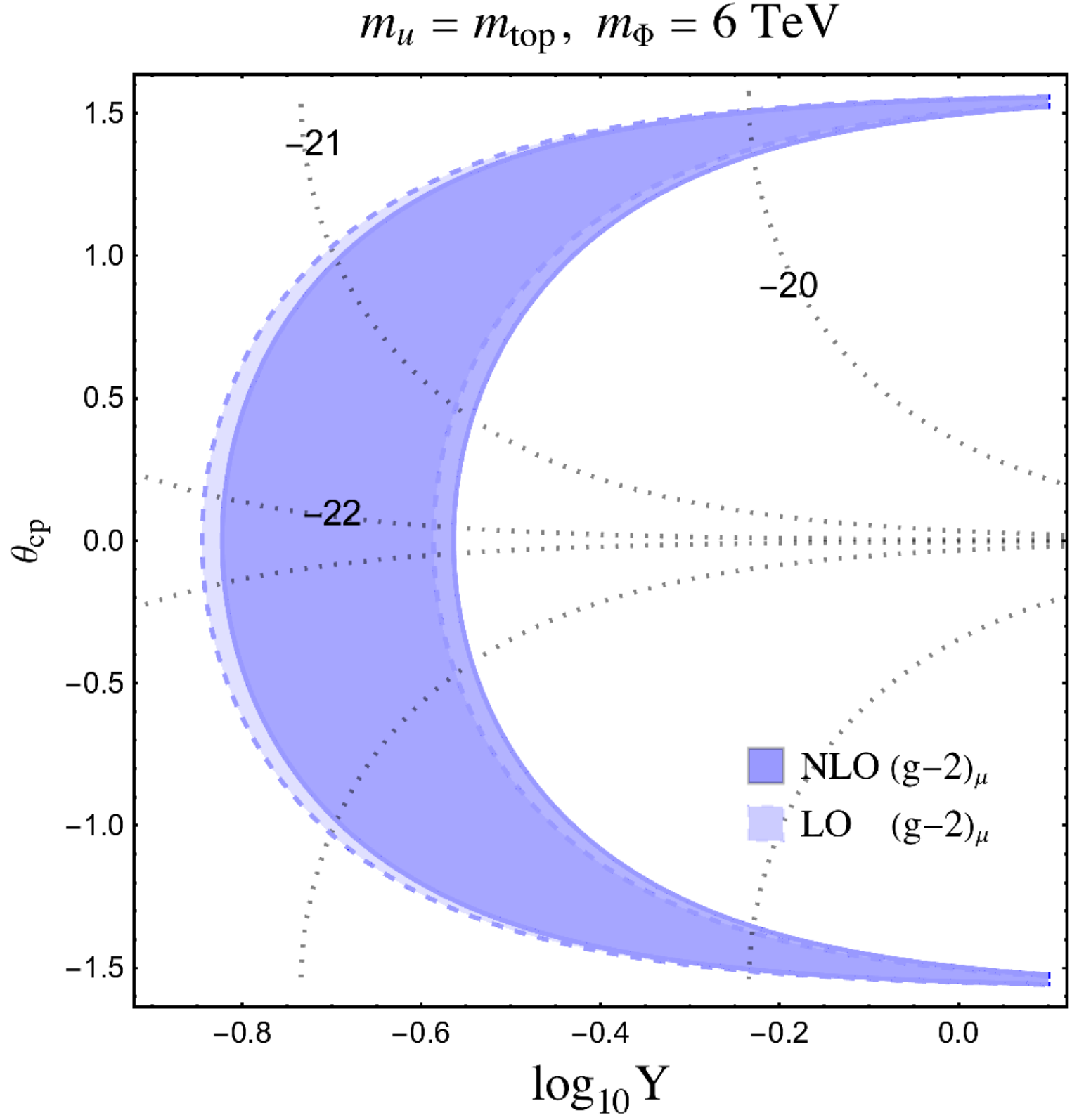}
		\caption{Phase and magnitude of the complex Yukawa coupling of the muon that match, in correspondence of the shaded regions, the measured $(g-2)_{\mu}$ $2\sigma$ interval. In both the panels, the dotted lines indicate the expected order of magnitude of the muon EDM ($e\cdot$cm). Left panel: a generic scenario with new physics at the TeV scale and for a critical value of $Q_u = -1/3$. The effect of NLO is clearly sizeable. Right panel: the same as in the left panel after identifying the fermion $u$ with the SM top quark. The two-loop contribution is negligible.}
	\label{LQ3plots3}
\end{figure}

For the scenario $m_{\Phi} \gg m_u\gg m_\ell$, the relative magnitude of the two-loop correction is an involved function of the available mass ratios and of the fermion charge $Q_u$. This is shown in the left panel of Fig.~\ref{LQ3plots1}, where we observe that the NLO is generally indistinguishable from the LO result for $m_u \lesssim 4$ TeV, barring a narrow region around the critical value $Q_u\simeq - 0.15$. For larger values of the fermion mass, instead, the correction is sizeable and affects the muon and electron AMM for a wider range of $u$ charges. The effect is analysed in the right panel of Fig.~\ref{LQ3plots1}, where we highlight the charge range yielding sizeable NLO corrections, as well as in the left panel of Fig.~\ref{LQ3plots2} where the NLO solutions are clearly distinguishable. The result is confirmed by the analysis of the correlation between muon AMM and EDM, shown in the left panel of Fig.~\ref{LQ3plots3}. Here we see that a fermion and a scalar states in the few TeV range still yield important NLO contributions of phenomenological interest.

\FloatBarrier

Differently, once the $u$ fermion is identified with the SM top quark (or with its Majorana conjugate), the two-loop contribution becomes clearly subdominant. The case is shown in right panels of Figs.~\ref{LQ3plots2} and~\ref{LQ3plots3}, where the NLO solutions inevitably overlap with the LO result on all of the considered parameter space. Two-loop corrections can then be safely neglected, for instance, within the $R_2$ leptoquark model ($u \equiv t, Q_u = 2/3$) as the positive $u$ charge prevents any NLO enhancement. As for the $S_1$ model, where $Q_u = -2/3$ ($u \equiv t^c, Q_u = -2/3$), we find instead that entering the NLO enhancement region requires values of the scalar field mass forbidden by collider searches.

\FloatBarrier

%-------------------------------------------------------------------------------
\section{Impact of precision $Z$ measurements}
\label{sec:RK}
%-------------------------------------------------------------------------------

On general grounds, accounting for the precision measurements of the $Z$ boson branching ratios requires the specification of an ultra-violet completion for the present framework. In fact, the considered new physics contributions to the muon and electron AMM exploit the simultaneous presence of both chiral couplings in the Yukawa structure presented in Tab.~\ref{ts}. While this is automatically the case if the $u$ fermion is identified with a SM quark, in more general scenarios it is therefore necessary to opportunely extend the Yukawa sector. A minimal setup is provided by the following Lagrangian:
\begin{align}
	\mathcal L &= \mathcal{L}_{SM} + \bar{f}(i\slashed{D}-m_f)f + \bar{\chi}(i\slashed{D}-m_\chi)\chi  + (D_\mu \Phi)^\dagger (D^\mu \Phi) - m^2_\Phi \Phi^\dagger \Phi + \\ \nn
	& +
	 \lambda_R \bar f_R^c l_R \Phi^\dag  + \lambda_{L} \bar \chi_L^c i \tau_2 L_L \Phi^\dag - k_1 \bar \chi_L i \tau_2 H^\dag f_R   - k_2 \bar \chi_R i \tau_2 H^\dag f_L    + \textrm{H.c.}.
	 \label{lagmod}
\end{align}
In the above equation $H$, $L_L$ and $l_R$ are the SM Higgs doublet, lepton doublet and right handed lepton, respectively. The scalar field $\Phi$ and the Dirac fermions $\chi$ and $f$ transform under the SM gauge group as detailed in Tab.~\ref{dofs}. Because the conservation of hypercharge requires
\begin{align}
	y_\Phi = y_f -1\nn, \qquad	y_\chi = y_f -\frac 12,
\end{align}
we identify the $u$ fermion of SLMs with the Majorana conjugated up component $\chi_u$ of $\chi$.  In fact, taking $\Phi$ as the corresponding scalar of SLMs, we correctly recover the relation among the QED charges $Q_\Phi- Q_{\chi_u} = Q_\Phi + Q_u = -1$ implied by Eq.~\eqref{yuk}. 

\begin{table}[htb!]
	\centering
	\begin{tabular}{c|c|c|c}
		\toprule
		Field: & $SU(3)_c$  & $SU(2)_L$&  $q_Y$ \\
		\midrule
		$\Phi$ &$ 3$ & $1$ &$ y_\Phi$ \\
		$f$ &$  3$ & $1$ &$ y_f$ \\ 
		$\chi$ &$ 3$ & $2$ &$ y_Q$ \\ 
		\bottomrule
	\end{tabular}
	\caption{Representation and charges of the considered degrees of freedom. }
	\label{dofs}
\end{table}

After the spontaneous symmetry breaking of weak interactions, the up component of $\chi$ and the fermion $f$ are mixed by new mass contributions quantified in 
\bea
 - \mathcal{L} \supset
(\bar f,  \bar \chi_u)_L \left( \begin{array}{cc}
m_f &  m_2 \\
m_1 & m_Q
\end{array}
\right)
\left( \begin{array}{c}
f \\ \chi_u
\end{array}
\right)_R,
\eea
where $\chi = (\chi_u, \chi_d)$ and $m_{1,2} = k_{1,2} \, v/\sqrt{2}$, $v=246$ GeV being the Higgs boson vacuum expectation value. The mass matrix is diagonalized through a biunitary transformation involving the mixing matrices $U_R$ and $U_L$, defined implicitly by
\bea
\left( \begin{array}{c}   f \\ \chi_u \end{array} \right)_R = \left( \begin{array}{cc}
\cos(\theta_R) & \sin(\theta_R) \\
-\sin(\theta_R) & \cos(\theta_R)
\end{array}
\right)
   \left( \begin{array}{c}   \psi_1 \\ \psi_2 \end{array} \right)_R,
\qquad
\left( \begin{array}{c}   f \\ \chi_u \end{array} \right)_L = \left( \begin{array}{cc}
	\cos(\theta_L) & \sin(\theta_L) \\
	-\sin(\theta_L) & \cos(\theta_L)
\end{array}
\right)
   \left( \begin{array}{c}   \psi_1 \\ \psi_2 \end{array} \right)_L.
\eea
In term of the mass eigenstates $\psi_{1,2}$, the Lagrangian consequently admits the following terms
\begin{equation}
	\mathcal L \supset 
\bar \psi_1 \left[\lambda_R \, \cos(\theta_L) P_R  - \lambda_L \,   \sin(\theta_R)    P_L\right] l  \Phi^\dag \\
+ 
\bar \psi_2 \left[\lambda_R \,   \sin(\theta_L)  P_R + \lambda_L \,     \cos(\theta_R)  P_L\right] l  \Phi^\dag 
\end{equation}
where $P_{L,R} = (1 \pm \gamma_5)/2$ are the usual chirality projectors. We remark that the spontaneous symmetry breaking is pivotal in recovering the structure of the SLM Lagrangian in Eq.~\eqref{yuk}.

With the above expression at hand, we checked the experimental constraints imposed by the lepton universality of $Z$ boson decays, finding that the bound mainly affects the axial current. However, in all the scenarios analyzed, we find it possible to avoid the constraint without impairing the AMM solution in a large part of the parameter space, which favours values of the mixing angles $\theta_L\simeq\theta_R$.

%-------------------------------------------------------------------------------
\section{Conclusions}\label{conclusions}

The planned and ongoing lepton experiments require an improvement in the precision of the corresponding theoretical predictions, needed to disentangle the possible effects of new physics. To this purpose, in the present paper we have investigated the structure of the lepton-photon vertex within extensions of the standard model that involve colored degrees of freedom coupled to leptons.

In order to detail the dominant corrections in a manner as general as possible, we have introduced the simplified leptoquark models: straightforward extensions of the standard model where the leptons interact with new colored and electrically charged degrees of freedom. The framework is therefore meant to reproduce the main phenomenological features of complete theories that propose such interactions, for instance leptoquark or grand unification models.  

In this first exploration of simplified leptoquark models we have considered the effects of an additional colored scalar particle, coupled to a standard model charged lepton and either a quark or a new colored fermionic field. With this simplified setup, which neglects subdominant weak interactions, we have computed the dominant two-loop corrections due to the new Yukawa and color interactions for three scenarios characterized by different mass hierarchies. The obtained expressions of the involved two-loop amplitudes constitute a first technical result of the analysis that improves on the  literature for the precision achieved. 

With these expressions at hand, we studied the three scenarios in isolation focusing in particular on the anomalous magnetic moment of the electron and muon, pending new experimental results regarding the latter. If the colored fermion involved in the interactions is not a quark of the Standard Models, we find regions of the parameter space where the two-loop contribution cannot be neglected. In fact, on these regions the one-loop corrections progressively vanishes as the electric charge of the colored fermion approaches a critical value, specific of the chosen mass hierarchy. On the contrary, we find that once the fermion is identified with a Standard Model quark, the two-loop contributions can safely be neglected. We argue that similar enhancements appear also in other leptonic precision observables because of the similar loop structures of the involved form factors.

%---------------------------------------------------------------

\section*{Acknowledgements}
This work was supported by the Estonian Research Council grants PRG356, PRG803, MOBTT86 and by the EU through the European Regional Development Fund
CoE program TK133 ``The Dark Side of the Universe". The authors thank M. Steinhauser for providing {\tt exp} and {\tt q2e}, as well as further guidance to their use.

\bibliographystyle{JHEP}
\bibliography{bib}

\end{document}